\documentclass[journal]{IEEEtran}

\let\labelindent\relax  
\usepackage{epstopdf}
\usepackage[draft]{hyperref} 
\usepackage{setspace}
\usepackage{amsmath}
\usepackage{amssymb}
\usepackage{stfloats}
\usepackage{amsfonts}
\usepackage{color}
\usepackage{graphicx}
\usepackage{subfigure}  

\usepackage{enumitem}
\usepackage[most]{tcolorbox}
\usepackage{cuted}
\usepackage[ruled,linesnumbered]{algorithm2e}  

\usepackage[flushleft]{threeparttable} 

\usepackage{upgreek}
\usepackage{bm}
\usepackage{nicefrac}

\usepackage{multirow}

\usepackage{cite}
\usepackage{mathtools}
\usepackage{stmaryrd}
\usepackage{algorithmic}
\usepackage[mathscr]{euscript}
\usepackage{array}
\usepackage{mathrsfs}
\usepackage{soul}
\usepackage{dsfont}

\newcommand{\myVec}[1]{{\boldsymbol{#1}}}
\newcommand{\myMat}[1]{{\boldsymbol{#1}}}
\newcommand{\mySet}[1]{\mathcal{#1}}

\newcommand{\myA}{\myMat{A}}

\definecolor{mypurple}{rgb}{0.910, 0.910, 0.969}
\definecolor{myblue}{rgb}{0.122, 0.435, 0.698}

\newtheorem{theorem}{Theorem}
\newtheorem{lemma}{Lemma}

\newtheorem{definition}{Definition}

\newcommand{\beq}{\begin{equation}}
\newcommand{\eeq}{\end{equation}}
\newcommand{\bea}{\begin{array}}
\newcommand{\ena}{\end{array}}
\newcommand{\bds}{\begin {itemize}}
\newcommand{\eds}{\end {itemize}}
\newcommand{\bdf}{\begin{definition}}
\newcommand{\blm}{\begin{lemma}}
\newcommand{\edf}{\end{definition}}
\newcommand{\elm}{\end{lemma}}
\newcommand{\bthm}{\begin{theorem}}
\newcommand{\ethm}{\end{theorem}}
\newcommand{\bprp}{\begin{prop}}
\newcommand{\eprp}{\end{prop}}
\newcommand{\bcl}{\begin{claim}}
\newcommand{\ecl}{\end{claim}}
\newcommand{\bcr}{\begin{coro}}
\newcommand{\ecr}{\end{coro}}
\newcommand{\bquest}{\begin{question}}
\newcommand{\equest}{\end{question}}

\newcommand{\larrow}{{\larrow}}

\def\urltilda{\kern -.15em\lower .7ex\hbox{\~{}}\kern .04em}

\IEEEoverridecommandlockouts

\begin{document}
\author{  Satish~Mulleti, {\it Member, IEEE}, Timur Zirtiloglu,  {\it Student Member, IEEE}, Arman Tan,  {\it Student Member, IEEE}, Rabia Tugce Yazicigil,  {\it Senior Member, IEEE}, Yonina C. Eldar, {\it Fellow, IEEE}
	\thanks{\scriptsize S. Mulleti is with the Department of Electrical Engineering, Indian Institute of Technology (IIT) Bombay, India, T. Zirtiloglu, A. Tan and R. T. Yazicigil are with the Electrical and Computer Engineering Department, Boston University, Boston, MA, USA, and Y. C. Eldar is with the Faculty of Math and Computer Science, Weizmann Institute of Science, Israel. Email: mulleti.satish@gmail.com, timurz@bu.edu, armantan@bu.edu, rty@bu.edu, yonina.eldar@weizmann.ac.il}
}
\title{Power-Efficient Sampling}
\markboth{the IEEE Signal Processing Magazine}%
{Shell \MakeLowercase{\textit{et al.}}: Bare Demo of IEEEtran.cls for Journals}
\maketitle

\begin{abstract}
Analog-to-digital converters (ADCs) facilitate the conversion of analog signals into a digital format. While the specific designs and settings of ADCs can vary depending on their applications, it is crucial in many modern applications to minimize their power consumption. The significance of low-power ADCs is particularly evident in fields like mobile and handheld devices reliant on battery operation. Key parameters of the ADCs that dictate the ADC's power are its sampling rate, dynamic range, and number of quantization bits. Typically, these parameters are required to be higher than a threshold value but can be reduced by using the structure of the signal and by leveraging preprocessing and the system application needs. In this review, we discuss four approaches relevant to a variety of applications.
\end{abstract}

\IEEEpeerreviewmaketitle

\section{Introduction}
 Real-world signals, including speech, audio, biomedical data, radar readings, communication transmissions, and more, inherently exist in an analog format. Transforming these signals into a digital representation is often a more efficient way to process, compress, store, and transmit them when considering factors like power efficiency, noise resistance, and algorithms' adaptability. Therefore, the digital conversion of analog signals is central to any application involving real-world signals.

To achieve this digital representation, analog-to-digital converters (ADCs) are employed. Given that ADCs are an essential component in many applications, their design and selection of parameters, such as the sampling rate, the number of bits for quantization, and the dynamic range (DR), are critical factors in effectively addressing the requirements of a particular application. For instance, when it comes to accurately reconstructing bandlimited signals from uniformly spaced samples, the ADC's sampling rate should meet or exceed the Nyquist rate, which is twice the maximum frequency component of the analog signal. In a similar vein, the dynamic range of the ADC should exceed that of the signal to prevent signal clipping during the sampling process, and a greater number of bits results in reduced quantization errors.

In emerging fields like wearable biosensors, wireless sensor networks, voice-activated interfaces (such as Amazon's Alexa, Apple's Siri, Microsoft's Cortana, and Google's Assistant), Internet of Things sensors, and more, the power consumption of ADCs holds great importance. Take, for example, wearable biosensors, which can monitor parameters like blood pressure, electrocardiogram readings, and epilepsy seizures. These devices often operate within stringent resource limitations, including power and space constraints, especially since many of them rely on battery power with finite energy reserves. In such applications, efficiently reducing power requirements for the sensor's ADCs is essential to ensure uninterrupted monitoring of the signals. Likewise, in any voice-activated interface, the voice detection sensors remain active continuously, underscoring the significance of minimizing the power demands of their ADCs. Furthermore, in applications that deal with wideband signals, the power consumption of the ADC is notably high due to the need for a high sampling rate, and this element constitutes a significant portion of the overall power requirements of the system.

In this review, we explore techniques and frameworks to lower the ADC power and the entire system power. A common theme is to enable analog preprocessing prior to sampling together with postprocessing on the signal in order to reduce the ADC power while still enabling signal recovery. The recovery may involve either reconstructing the original analog signals from the collected samples or estimating certain parameters. The power consumption of a conventional ADC, which captures uniformly spaced instantaneous samples of the analog signal, increases in tandem with the sampling rate, the number of bits used, and the dynamic range of the ADC. To address this challenge, we delve into theoretical lower limits on the power consumption of ADCs \cite{adc_power} and also examine a few figures-of-merit (FoMs) employed for assessing commercially available ADCs \cite{adc_power,walden_adc}. Once these relationships are established, various frameworks can be applied to mitigate the impact of these factors on power requirements.

We consider the following four approaches for power reduction: (a) \emph{Sub-Nyquist sampling}, where the signal's structure is used to reduce the sampling rate \cite{eldar_2015sampling}; (b) \emph{Modulo-ADC} where a low-dynamic range ADC preceded with a folding circuit is used to sample high-DR signals without clipping \cite{uls_tsp,bhandari2021unlimited, eyar_tsp, mulleti2022modulo}; (c) \emph{Asynchronous sampling} where time instants at which a signal crosses a set of thresholds are used as the discrete representation \cite{logan1977information,lazar2004perfect,gontier2014sampling,alexandru2019reconstructing,hila_tsp}; and (d) \emph{Low-bit quantization} methods in which the number of bits is reduced when the end task is to recover a few parameters of the signal \cite{shlezinger2019hardware,salamatian_2019}.

The majority of these frameworks comprise three primary components: (i) an analog pre-processing hardware or an analog front-end, (ii) a traditional low-power ADC, and (iii) a digital processing module. The pre-processing component alters the analog signal in a manner that allows the ADC to function at a reduced sampling rate, a lower bit rate, or a narrower dynamic range compared to an ADC lacking pre-processing. The digital processing segment encompasses algorithms that can potentially reverse the impact of the pre-processing stage and includes reconstruction or recovery algorithms. We detail these components' integrated design, considering theoretical and practical perspectives. In the theoretical realm, we present assurances for achieving perfect signal recovery and examine algorithms concerning critical parameters like the sampling rate, number of bits, and dynamic range. Subsequently, we explore the influence of these parameters on power consumption and hardware design. In addition, we also discuss existing hardware prototypes for these various approaches \cite{mishra_subnyquist_radar,mishali2011xampling, mishali2011bSPMag,bhandari2021unlimited,bhandari2022back, mulleti2023hardware,naaman2023hardware,gong_nir_hw,yazicigilSPM}.

In the following section, we discuss bounds and FoMs, which will be used to assess the power-saving aspects of the considered frameworks.

\begin{figure}
     \centering
     \includegraphics[width = 2in]{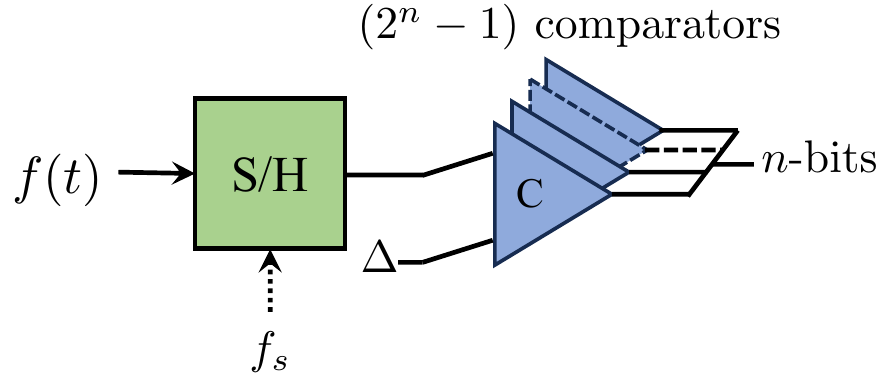}
     \caption{A conventional flash-type ADC with a sample and hold circuit operating at a sampling rate $f_s$, and a quantizer consisting of $2^n$ comparators where $n$ is the number of bits per sample. We assume that the analog signal $f(t)$ is bounded: $|f(t)|\leq A$. The quantization step size is $\Delta = A/(2^{n-1})$.}
     \label{fig:conventiona_adc}
 \end{figure}

\begin{tcolorbox}[float*=t,
    width=2\linewidth,
	toprule = 0mm,
	bottomrule = 0mm,
	leftrule = 0mm,
	rightrule = 0mm,
	arc = 0mm,
	colframe = myblue,
	colback = mypurple,
	fonttitle = \sffamily\bfseries\large,
	title = Bounds on Power Consumption of a Conventional ADC]	

    To compute lower bounds on the power, we consider two major components of a conventional ADC: (a) a sample and hold (S/H) circuit that samples and retains the value of the analog signal for quantization; (b) a quantizer that converts the held value from S/H to a bit stream (See Fig.~\ref{fig:conventiona_adc}) using comparators. The total power consumption is the sum of that of the S/H circuit and the comparators. To derive the total power, we assume that the capacitors used in the S/H and quantizers are chosen to be large enough such that the thermal noise is less than or equal to the quantization error \cite{adc_power}. For deriving the quantization error, we assume that the input signal $f(t)$ lies in the amplitude range $[-A, A]$ and the quantizer has $n$ bits. With these assumptions, the power consumption in the S/H is given as
    \begin{align}
        P_S = 24kT f_s 2^{2n},
        \label{eq:sh_power}
    \end{align}
        where $k$ is Boltzmann's constant and $T$ is the absolute temperature in degrees Kelvin. In the quantizer part, an expression for power consumption in each comparator is derived as
         \begin{align}
        P_C = 4n \, \ln2 \,V_{\text{eff}} C_C f_s  A  ,
        \label{eq:flash_power}
    \end{align}
      where $C_C$ is the load capacitance of the latch comparator. The quantity  $V_{\text{eff}}$ is a parameter of the comparator and is defined as $V_{\text{eff}} = g_m \, I_D$ where $I_D$ is the supply current of the comparator, and $g_m$ is the minimum transconductance of the comparator to make a decision within the sampling period. Hence, the total power of a conventional flash ADC is
      \begin{align}
          P = P_S + (2^n-1)P_C = 24kT f_s 2^{2n}+ 4(2^n-1)n \, \ln2 \,V_{\text{eff}} C_C f_s  A.
          \label{eq:adc_power}
      \end{align}
\label{Box:adc_power}
\end{tcolorbox}	

\section{Power Consumption of an ADC and FoMs }
\label{sec:traditionalADC}
We first discuss bounds on the power consumption of a conventional ADC, which are used throughout the paper to compare different ADCs. The details are given in the box \emph{Bounds on Power Consumption of a Conventional ADC}. We refer the readers to \cite{adc_power} and references therein for details on these derivations.

From \eqref{eq:adc_power}, we infer that the power consumption increases linearly with the sampling rate and exponentially with the number of bits. These observations are based on theoretical analysis, and in practice, the power dependency on the sampling rate and number of bits could be different. To understand the practical aspect, we consider the surveys considered in \cite{walden_adc} and a recent one by Murmann \cite{adc_survey}. These reports compare various ADCs in terms of different FoMs. As an example, Walden's FoM for a conventional ADC is given as \cite{walden_adc}
\begin{align}
        \text{FOM}_{\text{W}} = \frac{P}{2^{\text{ENOB}} \, f_s},
	\label{FOM:walden}
\end{align}
where $P$ is the power consumption of the ADC, $f_s$ is the sampling rate, and ENOB is the effective number of bits. An empirical study of the available ADCs in \cite{adc_survey} shows that the $\text{FOM}_{\text{W}}$ does not change as $f_s$ changes for $f_s \leq 100$ MHz. This implies that $P$ is proportional to $f_s$. However, for $f_s > 100$ MHz, the behavior of $\text{FOM}_{\text{W}}$ as a function of $f_s$ reveals that the increase in power consumption due to the sampling frequency exceeds linear increment, which means that the actual power saving of sub-Nyquist is more than the theoretical power saving. Further, the reviews reveal that, in commercially available ADCs, the sampling rate and number of bits are typically not independent. Specifically, the number of bits decreases with an increased sampling rate owing to design constraints. 

In the aforementioned analysis, we do not observe an explicit dependency of power consumption on the DR. Rather, power consumption is an implicit function of DR, as shown next. Consider two ADCs with dynamic ranges $[-\lambda_1, \lambda_1]$ and $[-\lambda_2, \lambda_2]$ respectively, where $\lambda_1>\lambda_2$. Let the resolutions of the two ADCs be $N_1$ and $N_2$ bits, respectively. Then, the quantization noise powers are given as $\frac{1}{12}\left(\frac{\lambda_1}{2^{N_1-1}} \right)^2$ and $\frac{1}{12}\left(\frac{\lambda_2}{2^{N_2-1}} \right)^2$. To keep the same quantization error for both the ADCs, the number of bits should satisfy the equality $N_1 = N_2 + \frac{\lambda_1}{\lambda_2} \log 2$. Hence, for a high-DR ADC, one must choose a larger number of bits to keep the same quantization error level. Following \eqref{eq:adc_power}, this results in higher power consumption. 

In summary, the power consumption of an ADC increases with the sampling rate, number of bits, and DR. In the following, we discuss how to reduce these quantities and eventually the power consumption without compromising the quality of the reconstruction or task. In general, these quantities are reduced by analog preprocessing steps, which could consist of either single-channel or multiple-channel front-end circuits. These front ends could be realized using off-the-shelf components or custom-designed analog and radio-frequency circuits with low-power consumption by co-designing digital signal processing and analog preprocessing. Hence, the overall power consumption of the circuit can be made lower than the corresponding conventional circuit.


\section{Sub-Nyquist Sampling}
\label{sec:sub-nyquist}
In most practical applications, analog signals are either assumed bandlimited or maximally bandlimited. In the latter case, one assumes that most (e.g., $99 \%$) of the signal's energy lies in a frequency range known as the signal's effective bandwidth. These signals are sampled according to the well-known Shannon-Nyquist sampling framework \cite{eldar_2015sampling}. In this framework, the signals are first passed through an anti-aliasing filter to project the signal to a bandlimited space and then sampled uniformly at the Nyquist rate, which is twice the maximum frequency content of the signal. Since the power consumption of ADCs increases with the sampling rate, the Shannon-Nyquist framework leads to higher power consumption for wideband applications such as ultra-wideband radar, cognitive radio, and wideband communications, where the bandwidth of the signals is on the order of GHz.

A solution to this problem is to explore signal structures beyond bandlimitedness and reduce the sampling rate below the Nyquist rate (sub-Nyquist sampling) by using the signal structure or model. Many signals encountered in practice have sparse representations either in the time or frequency domains, which leads to sub-Nyquist sampling. Several review articles on sub-Nyquist sampling exist from both theoretical aspects \cite{mishali2011bSPMag, mishali2011SPMag, eldar2009SPMag, eldar_2015sampling} and from practical considerations \cite{yazicigilSPM,timur_power}. Hence, in this review, we keep discussion on the sub-Nyquist theory short and focus on the power savings that can be obtained through this approach. In particular, we consider two signal models, finite-rate-of-innovation (FRI) signals \cite{vetterli,fri_strang,bluspmag, eldar_sos, mulleti_kernal} and multiband signals \cite{lin_ppv_98,herley_wong,mishali_2009,mishali_2010,mishali2011xampling}, and discuss their power efficiency aspects. 

\begin{figure}
    \centering
    \includegraphics[width = 3.5 in]{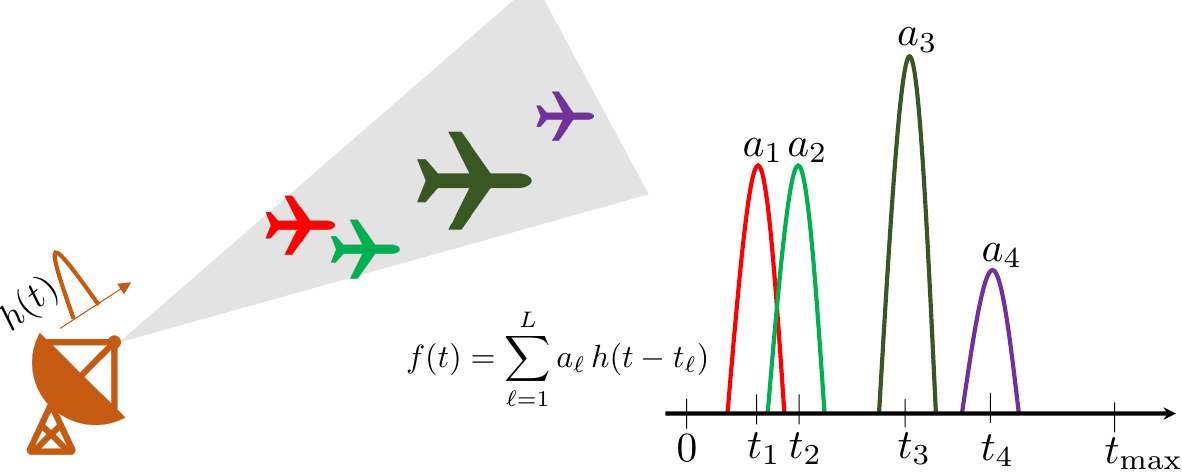}
    \caption{Application of FRI signals in radar imaging: A known transmit pulse $h(t)$ is reflected from sparsely located targets; the received signal is modeled as in \eqref{eq:fri} where time delays and amplitudes parameterize the signal.}
    \label{fig:radar}
\end{figure}
\begin{figure*}[!t]
    \centering
    \includegraphics[width = 4.5 in]{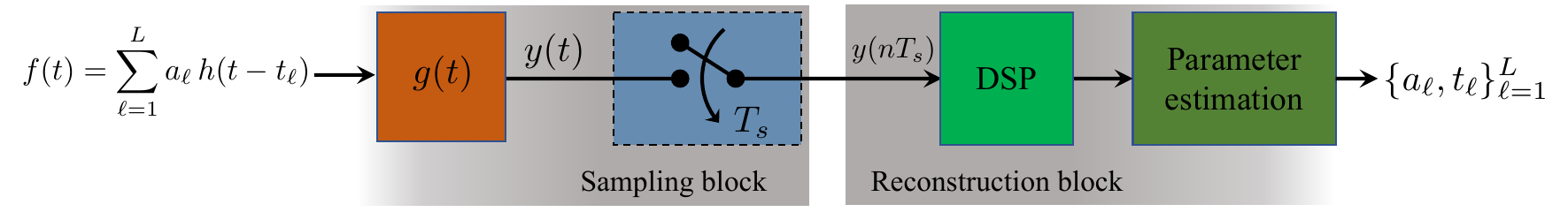}
    \caption{A schematic of FRI sampling and reconstruction framework}
    \label{fig:fri_framework}
\end{figure*}

\subsection{Finite-Rate-of-Innovation Signals}
In many applications, the signals can be specified by a finite number of parameters per unit time interval. As the rate of innovation (RoI) or the number of parameters specifying the signal per unit time interval is finite, these classes of signals are known as FRI signals. These signals can be sampled at their RoI; if the RoI is lower than the Nyquist rate, then it is sub-Nyquist sampling.

Commonly used FRI signals are streams of known pulses representing signals in time-of-flight applications such as radar, ultrasound, optical coherence tomography, and more.
An example is depicted in Fig.~\ref{fig:radar} where a known transmit pulse $h(t)$ reflects from sparsely located $L$ point targets. The received signal is given by
\begin{align}
    f(t) = \sum_{\ell=1}^L a_\ell \, h(t-t_\ell),
    \label{eq:fri}
\end{align}
where the amplitude $a_\ell$ and delay $t_\ell \in (0, t_{\max}]$ denote size and location of the $\ell$-th target, and $t_{\max}$ is maximum time delay. The signal $f(t)$ is specified by $2L$ parameters: $\{a_\ell, t_\ell\}_{\ell=1}^L$ and hence is an FRI signal. The Nyquist rate of these signals depends on the essential bandwidth of the pulse $h(t)$, which is generally very high due to its short time duration. However, due to the FRI nature, the parameters can be determined from sub-Nyquist samples. 

In Fig.~\ref{fig:fri_framework}, we depict a frequently employed FRI sampling and reconstruction framework consisting of a sampling and a reconstruction block. The sampling kernel $g(t)$ acts like an anti-aliasing filter in bandlimited sampling and plays a key role in the sampling block. The kernel removes additional information from the FRI signal and ensures that the filtered output $y(t) = (f*g)(t)$ is in a form such that the FRI parameters can be uniquely determined from low-rate samples of $y(nT_s)$. The reconstruction block, typically, consists of a digital signal processing (DSP) unit that linearly combines samples of $y(nT_s)$ and a parameter estimation method that determines the FRI parameters $\{a_\ell, t_\ell \}_{\ell=1}^L$. 

The sampling and reconstruction blocks are interdependent and must be designed in unison. To elaborate on this,  we consider a compactly supported sum-of-sincs (SoS) kernel with impulse response given as \cite{eldar_sos}
\begin{align}
    g(t) = \text{rect}\left(\frac{t}{T_g} \right)\sum_{k=-K}^K e^{\mathrm{j}k\omega_0 t}, \label{eq:sos}
\end{align}
where $\omega_0 >0$ and $\left(\frac{t}{T_g} \right) = 1$ for $t \in [0, T_g]$ and zero elsewhere. If the support of the filter $T_g$ is sufficiently larger than that of the signal $f(t)$, then a part of the filtered signal can be written as 
\begin{align}
    y(t) = \sum_{k=-K}^K F(k\omega_0) e^{-\mathrm{j}k\omega_0 t}, \label{eq:sos_op}
\end{align}
where $F(\omega)$ is the continuous-time Fourier transform (CTFT) of $f(t)$ (cf. \cite{eldar_sos, mulleti_kernal} for details). By choosing  $K\geq L$ and measuring $2K+1$ or more uniform samples $y(nT_s)$, one can determine the Fourier samples $F(k\omega_0)$ uniquely provided that $\omega_s = \frac{2\pi}{T_s} \geq (2K+1)\omega_0$. 

\begin{figure*}[!t]
	\begin{center}
		\begin{tabular}{cc}
			\subfigure[An FRI pulse $h(t)$]{\includegraphics[width=3in]{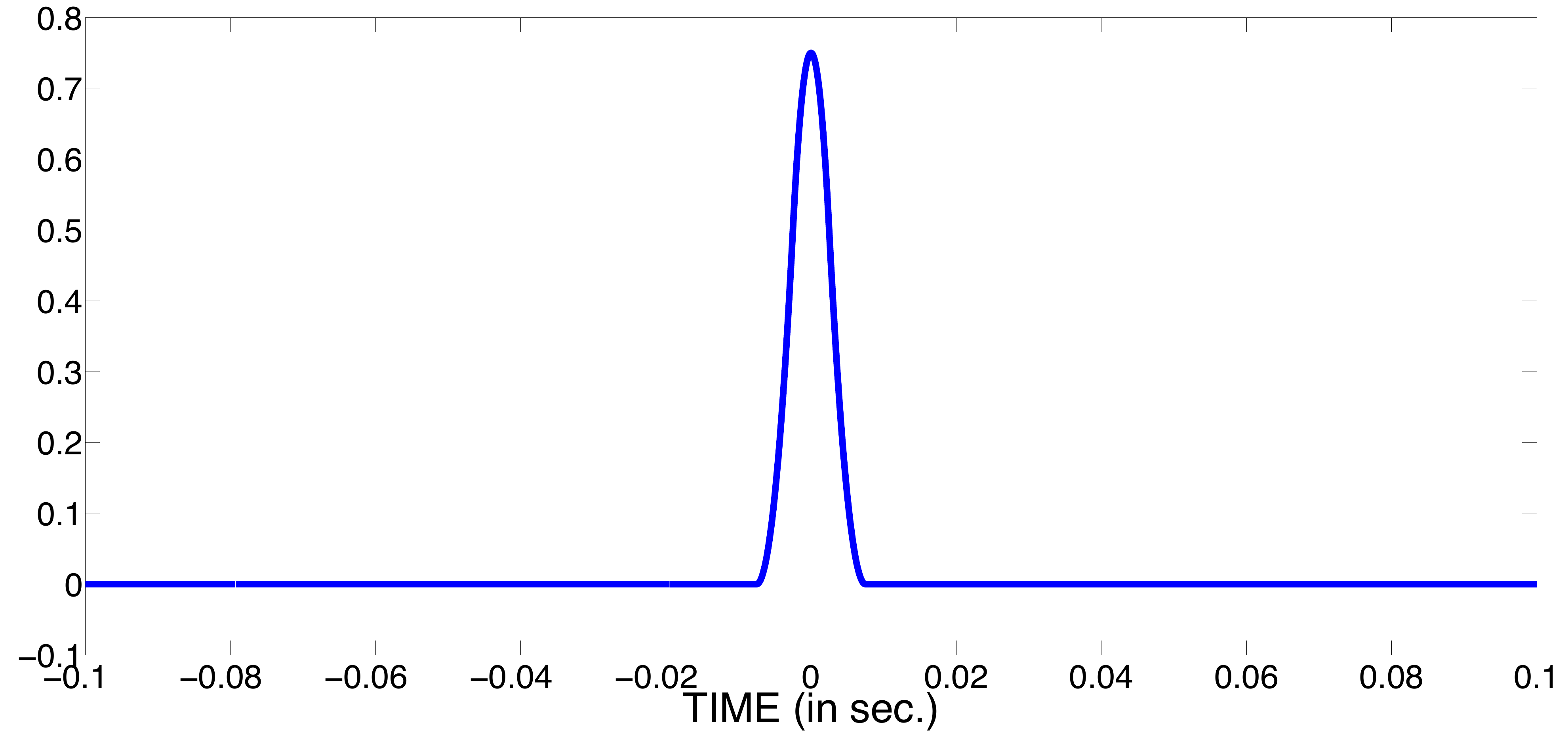}\label{fig:FRIpulse}} 
			\subfigure[The magnitude spectrum of $h(t)$]{\includegraphics[width=3in]{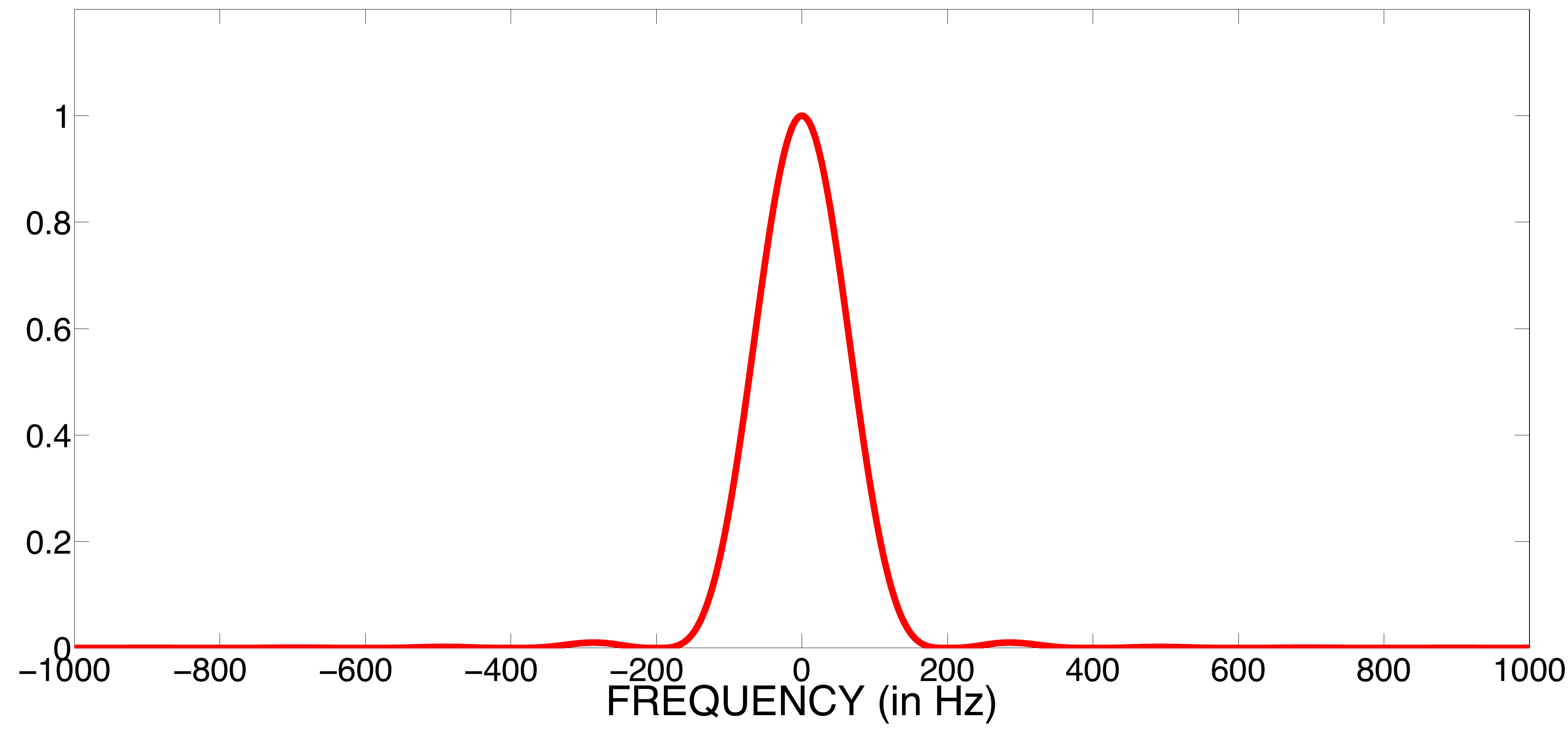}\label{fig:FRIpulse_ft}} \\
          \subfigure[FRI signal $f(t) = \sum_{\ell=1}^L a_\ell h(t-t_{\ell})$, $L=5$]{\includegraphics[width=3in]{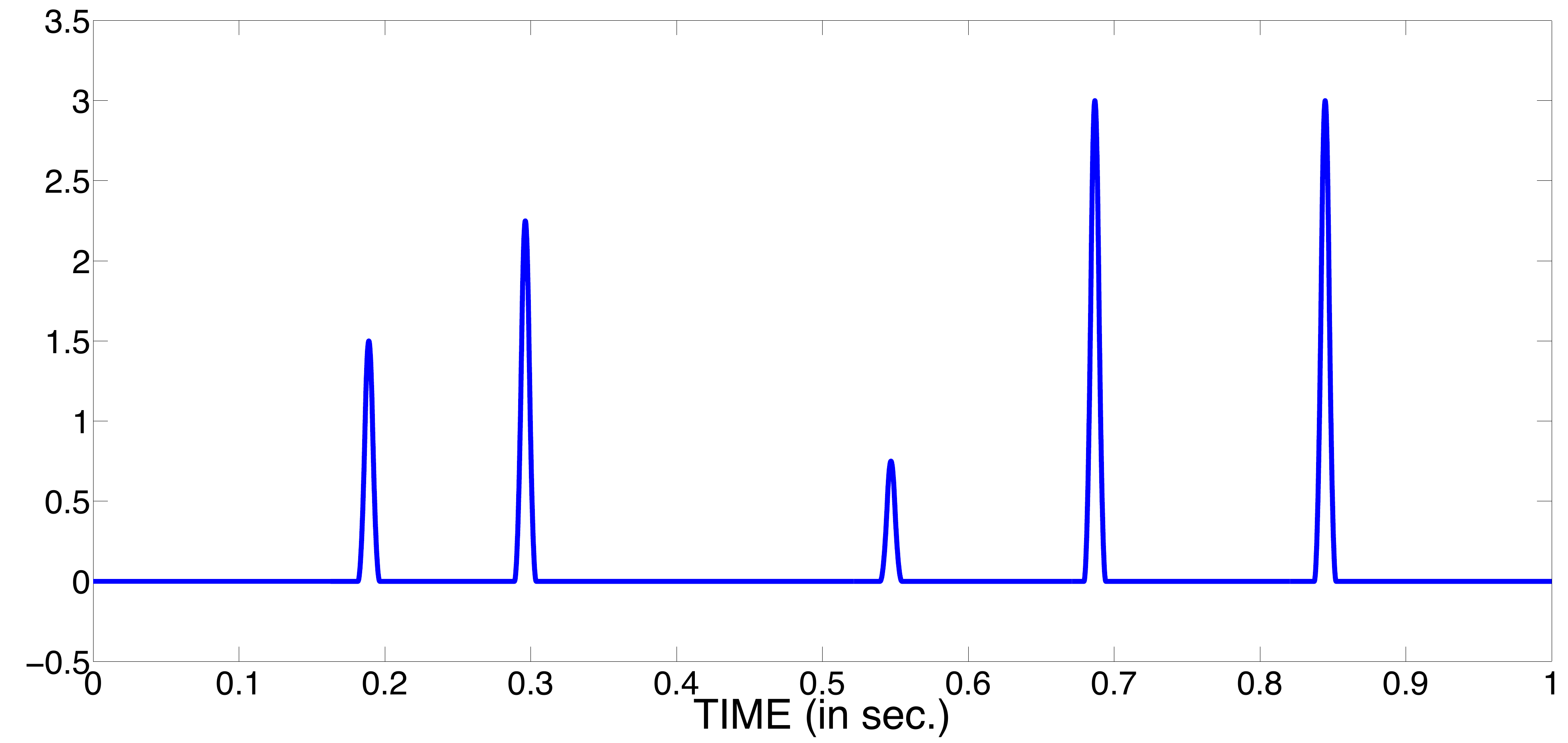}}\label{fig:FRIex}
			\subfigure[Magnitude spectrum of $f(t)$]{\includegraphics[width=3in]{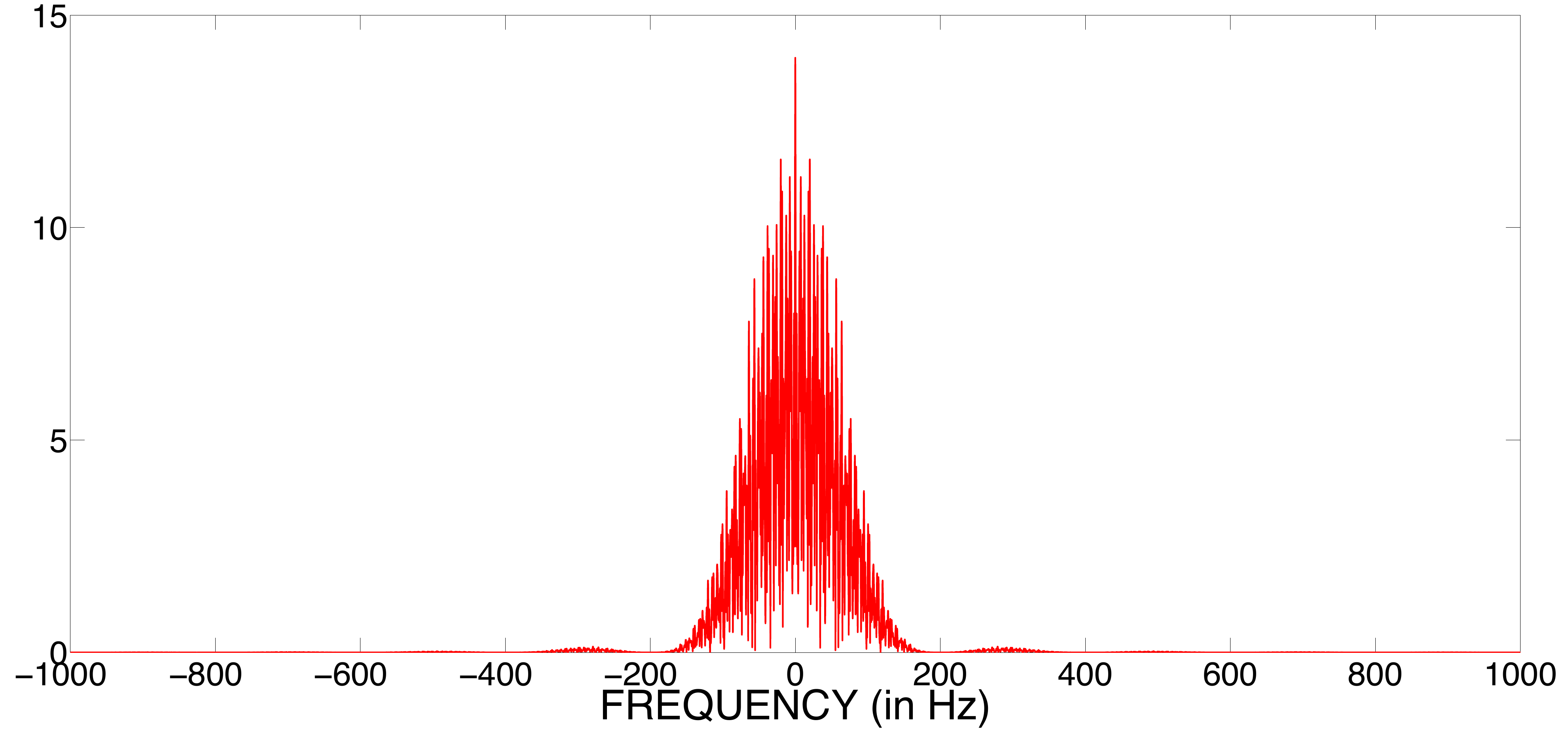}\label{fig:FRIexFT}}
		\end{tabular}
		\caption{An illustration of sub-Nyquist sampling of FRI signals: (a) $h(t)$ is a time-limited pulse with its magnitude spectrum $|H(\omega)|$ shown in (b). In (c), an FRI signal is shown consisting of $L=5$ delayed copies of $h(t)$ with $t_{\max} = 1$ sec. The magnitude spectrum of $f(t)$ shown in (d) shows that the Nyquist sampling rate is 400 Hz; however, the FRI-based framework requires sampling at 11 Hz.}
		\label{fig:fri_nyquist}
	\end{center}
\end{figure*}
From the Fourier samples, one can construct the following sequence
\begin{align}
    S(k\omega_0) = \frac{F(k\omega_0)}{H(k\omega_0)} = \sum_{\ell=1}^L a_\ell e^{\mathrm{j}k\omega_0 t_\ell}, \quad k = -K,\cdots, K
    \label{eq:S}
\end{align}
where $F(\omega)$ and $H(\omega)$ are CTFTs of $f(t)$ and $h(t)$, respectively, and $\omega_0$ is sampling interval in the Fourier domain. The sequence $S(k\omega_0)$ consists of a linear combination of complex exponentials with discrete frequencies $\{\omega_0 t_\ell\}_{\ell=1}^L$. 
The parameters $\{a_\ell, t_\ell \}_{\ell=1}^L$ can be uniquely determined from the sequence $S(k\omega_0)$ by using high-resolution spectral estimation methods \cite[Ch. 4]{stoica} provided that $\omega_0 t_{\max}<2\pi$ and $K\geq L$. In this approach, we assume that $H(k\omega_0) \neq 0, k = -K, \cdots, K$, which can always be ensured in practice as $h(t)$ has a narrow duration in time and therefore large bandwidth.

Similar spectral-estimation-based reconstruction can also be achieved by using a lowpass filter or a Gaussian kernel \cite{vetterli}. However, these kernels have infinite support and are, hence, impractical to realize. Other choices of compactly supported sampling kernels include polynomial and exponential-generating kernels \cite{blu_moms, fri_strang}. All the kernels have a common working principle: Spread the information of the FRI pulses such that each sample of the filtered signal contains sufficient information about the amplitude and time delays of all the pulses.

\begin{figure}
    \centering
    \includegraphics[width = 2.5 in]{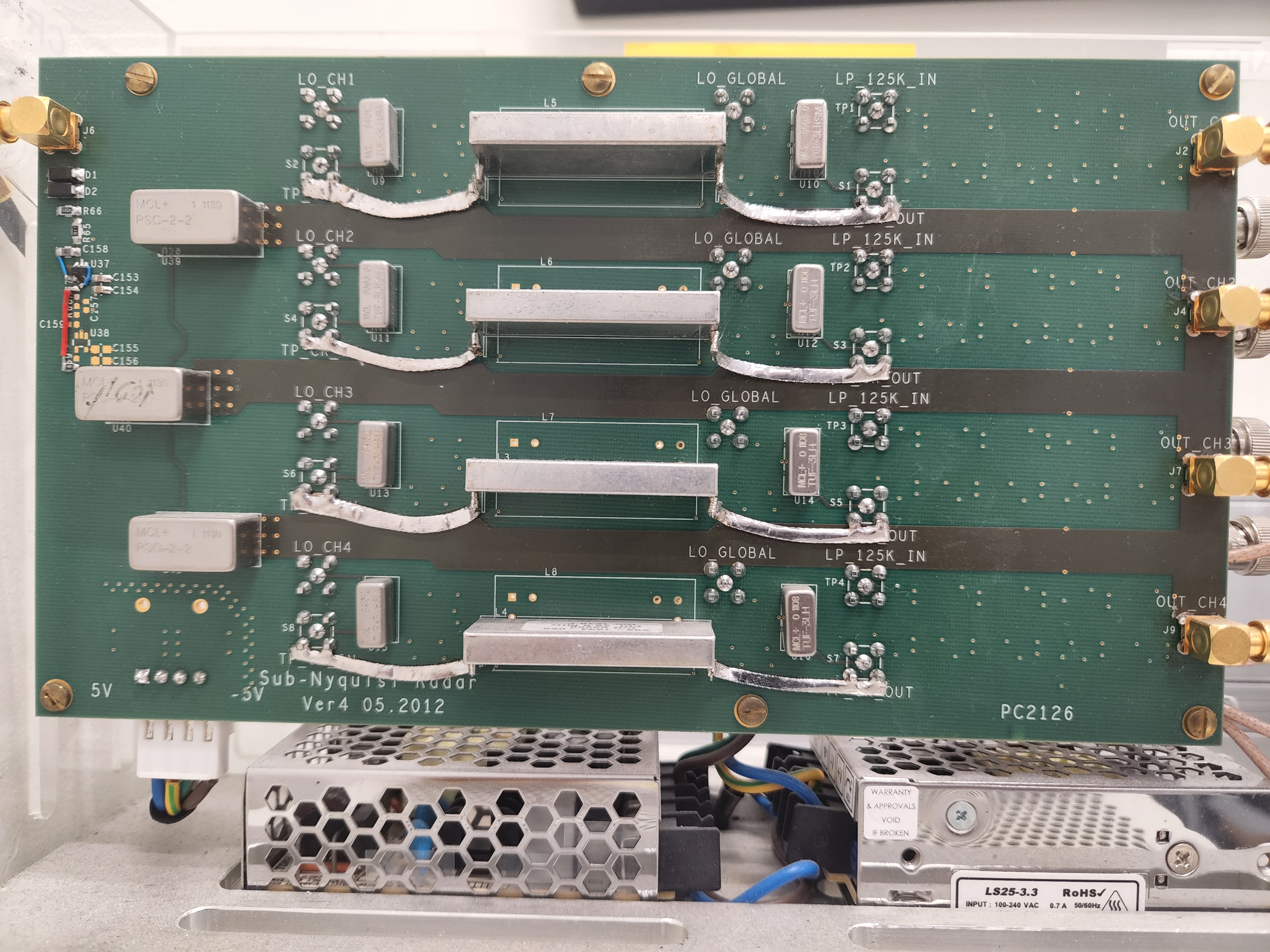}
    \caption{A hardware prototype of sub-Nyquist radar \cite{mishra_subnyquist_radar}.}
    \label{fig:sn_radar_hw}
\end{figure}
From the inequalities $\omega_s = \frac{2\pi}{T_s} \geq (2K+1)\omega_0$ and $|\omega_0 t_{\max}|<2\pi$, we infer that the lowest possible sampling rate is $(2L+1)\frac{2\pi}{t_{\max}}$. The sampling rate is independent of the bandwidth of the FRI signal and only a function of the number of pulses and maximum time delay. Indeed, this results in sub-Nyquist sampling, as illustrated in the subsequent example. In Fig.~\ref{fig:fri_nyquist}, we show a particular example of an FRI signal where $h(t)$ is a time-limited pulse constructed from a time-scaled, third-order B-spline function. The FRI signal consists of $L=5$ pulses with a maximum time delay of one second. The magnitude spectra of the pulse $h(t)$ and $f(t)$ show that the essential bandwidth is 200 Hz, which shows that the Nyquist rate is 400 Hz. However, by using the FRI nature of the signal, we know that it can be sampled at a minimum rate of $2L+1$ or 11 Hz. Hence, there is a reduction in the sampling rate by $~36$ times in this example, and consequently, power saving will be of the same order. However, one may not achieve such a large sampling rate reduction in practice due to noise and other practical limitations such as non-ideal sampling kernels. For example, hardware prototypes for sub-Nyquist radar systems, where the sampling and reconstruction are based on the FRI principle, are discussed in \cite{mishra_subnyquist_radar}. A hardware board is shown in Fig.~\ref{fig:sn_radar_hw}. The hardware prototypes demonstrate that using the received signals' FRI nature, the radar targets are estimated from the samples measured at $1/30$ of the Nyquist rate. As the sampling rate is reduced by a factor of 30 in this case, following \eqref{eq:adc_power}, power consumption is reduced by at least a factor of 30. If the operating frequencies are in the range of a few hundred MHz, then the power consumption goes down further, as discussed earlier, with the help of $\text{FOM}_{W}$.

\begin{figure}
    \centering
    \includegraphics[width = 3 in]{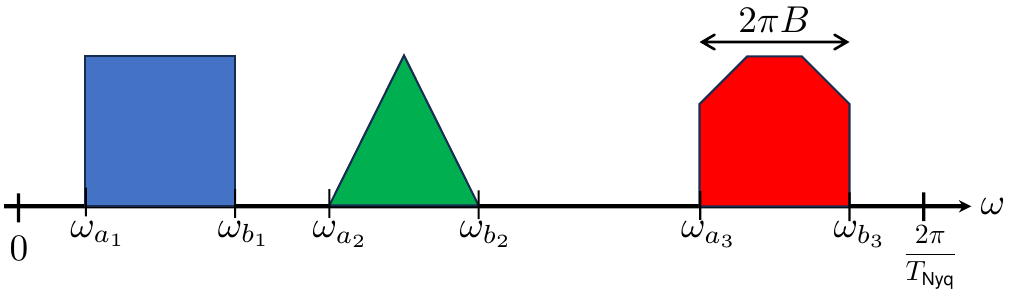}
    \caption{Spectrum of a multiband signal consisting of three disjoint bands. The maximum frequency of the signal is $2\pi/T_{\text{Nyq}}$, and the Nyquist rate is $1/T_{\text{Nyq}}$ Hz. In this example, for simplicity, we consider a signal with a positive spectrum, whereas signals in practice can have both positive and negative parts.   }
    \label{fig:multiband}
\end{figure}

\begin{figure}
    \centering
    \includegraphics[width = 3.5in]{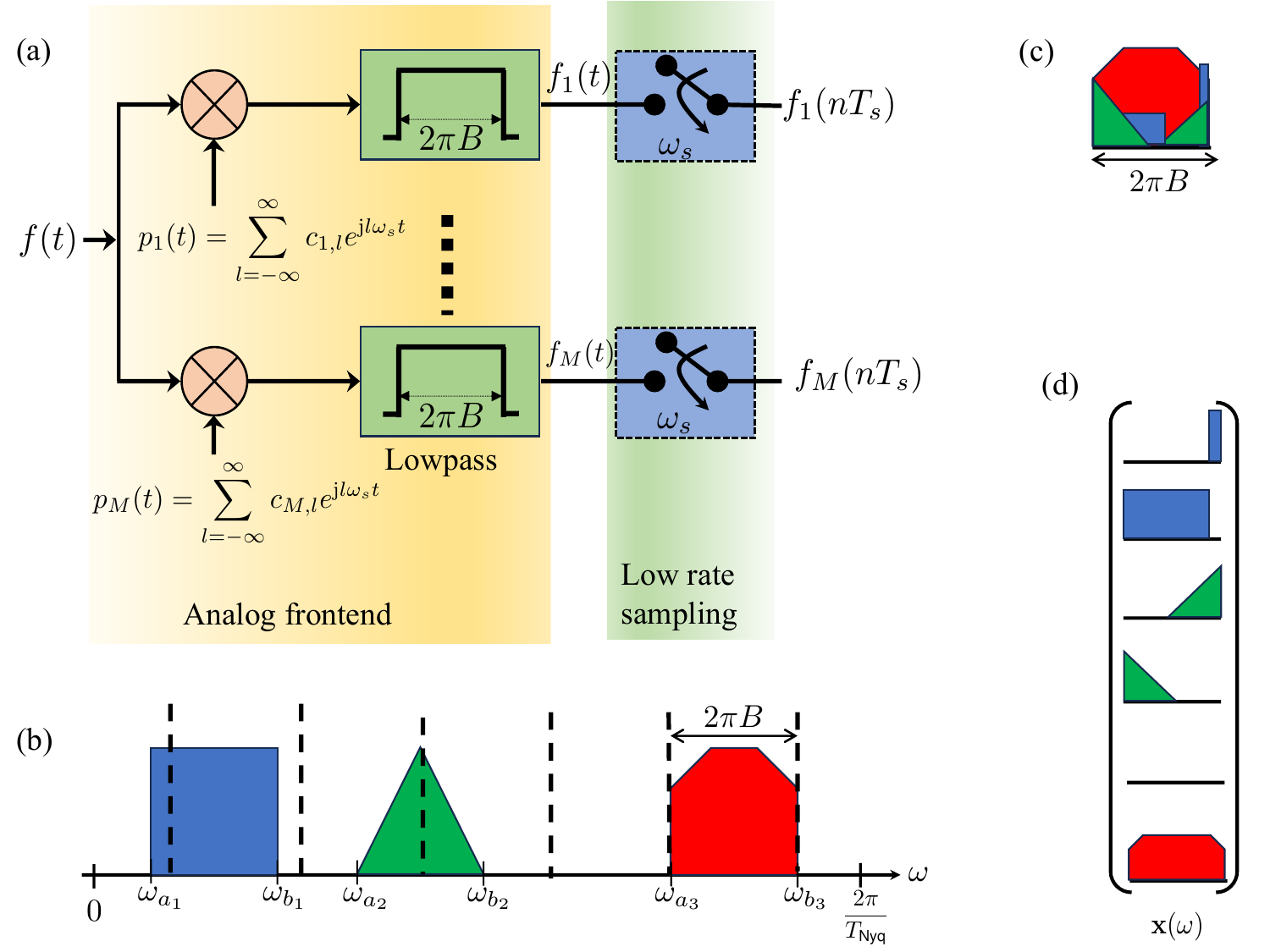}
    \caption{An $M$-channel MWC-based mixing and sub-Nyquist framework for multiband signals \cite{mishali_2009,mishali_2010,mishali2011xampling}.}
    \label{fig:mwc}
\end{figure}
 
\subsection{Sub-Nyquist Sampling of Multiband Signals}
To simultaneously communicate several bandlimited signals over a common channel, signals are modulated to different carrier frequencies,  as shown in Fig.~\ref{fig:multiband}. Following the Shannon-Nyquist sampling framework, such multiband signals can be sampled at twice the maximum frequency bandwidth of the signal. However, as shown in the example in Fig.~\ref{fig:multiband}, the signal's spectrum is zero over several intervals, and the information-bearing part resides in a few bands. Mathematically, consider a set of $N$ disjoint frequency intervals $\mathcal{M} = \cup_{n=1}^N[\omega_{b_n} - \omega_{a_n}] \subset [0, 2\pi/T_{\text{Nyq}}]$ where $\omega_{a_n}$ and $\omega_{b_n}$ denotes lower and upper band edges of the $n$-th band. Then $\mathcal{B}_{\mathcal{M}}$ denotes a set of signals whose spectrum is restricted to the interval $\mathcal{M}$. Specifically, if $f(t) \in \mathcal{B}_{\mathcal{M}}$ then $F(\omega) = 0, \omega \not\in \mathcal{M}$. The Nyquist rate of the signals in this class is $1/T_{\text{Nyq}}$, which could be much larger than the spectral support $|\mathcal{M}|$.

By observing the sparsity of the signal's spectrum, Landau \cite{landau} showed that a multiband signal is uniquely identifiable from its stable samples measured at a rate equal to its spectral support (referred to as Landau rate). Hence for signals in $\mathcal{B}_{\mathcal{M}}$, sampling at a rate $|\mathcal{M}|$ is sufficient. Periodic non-uniform sampling and multi-coset sampling patterns can be used for sampling and reconstructing multi-band signals \cite{lin_ppv_98, herley_wong}. Most of these approaches show that perfect reconstruction can be achieved by sampling the signals at the Landau rate, provided that the band edges $\{\omega_{a_n}, \omega_{b_n}\}_{n=1}^N$ are known.

A blind sampling and reconstruction framework was presented in a series of papers 
\cite{mishali_2009,mishali_2010,mishali2011xampling}, where the authors proved that the minimum sampling rate is twice Landau's rate and not more than the Nyquist rate in the blind setup. In these blind-multiband works, the authors assumed that $\omega_{b_n} - \omega_{a_n} = 2\pi B$ and $B$ and $N$ are known. Then, the set of multiband signals can be uniquely identifiable from stable samples measured at a rate $\min \{2NB, 1/T_{\text{Nyq}}\}$. An oversampling by a factor of two is the price paid for blindness. The signal reconstruction algorithms in these works first estimate the unknown spectral supports and then use a conventional algorithm to reconstruct the signals from the samples. For support recovery, the algorithms rely on a similar principle as in FRI sampling: to spread or mix the information so that the desired information is determined from fewer samples relying on notions from compressed sensing.

For mixing the information, the authors proposed two multichannel frameworks that are based on multi-coset sampling \cite{ herley_wong}, quadrature analog-to-information converter (QAIC) \cite{JSSCYaz2015, RFICYaz2016}, and the modulated-wideband converter (MWC) \cite{mishali_2010, mishali2011xampling,mishali2011bSPMag}. The principle idea of mixing is based on aliasing of the spectrum. Reconstruction is achieved by applying sparse-recovery methods from compressive sensing \cite{eldar_cs_book} as discussed next. For an $M$-channel sampling framework, an MWC-based mixing and sampling framework is shown in Fig.~\ref{fig:mwc}. The mixing is performed by multiplying $f(t)$ with a set of $M$ periodic signals $p_m(t) = \sum_{l = -\infty}^{\infty}c_{m, l} \, e^{\mathrm{j}\omega_s t}, m = 1, \cdots, M$, where $\omega_s \geq 2\pi B$ and $c_{m, l}$s are Fourier coefficients of the periodic signals. This results in modulation in the frequency domain and, as a result, the lowpass spectrum of the product $f(t) p_m(t)$ is a linear combination of all the bands as shown in Fig.~\ref{fig:mwc}(c). The mixed spectrum is captured from the sampled signal by passing the product $f(t) p_m(t)$ through a lowpass filter and then sample uniformly at its Nyquist rate. The discrete-time Fourier transform of the samples is related to the spectrum of the continuous-time multiband signal as \cite{mishali_2009, mishali_2010}
\begin{align}
    F_m(e^{\mathrm{j}\omega T_s}) = \sum_{\ell = -L}^L c_{m, \ell}\, F(\omega - \ell \omega_s), \quad \omega\in [-\omega_s/2, \omega_s/2], \label{eq:mwc}
\end{align}
where $\omega_s = \frac{2\pi}{T_s} < \omega_{\text{Nyq}}$ is the sampling rate in rad/sec. The summation in \eqref{eq:mwc} is due to aliasing because of sub-Nyquist sampling, and its finiteness is a consequence of the bandlimitedness of $f(t)$. Re-writing \eqref{eq:mwc} in matrix-vector notation, we have
\begin{align}
    \mathbf{y}(\omega) = \mathbf{A}\mathbf{x}(\omega) \in \mathbb{C}^M,  \quad \omega\in [-\omega_s/2, \omega_s/2], \label{eq:mwc2}
\end{align}
where $\mathbf{A}_{m,\ell} = c_{m,\ell}$, $\mathbf{y}(\omega) = [F_1(e^{\mathrm{j}\omega T_s}),\cdots, F_M(e^{\mathrm{j}\omega T_s})]^{\mathrm{T}}$, and $\mathbf{x}(\omega) = [F(\omega - L \omega_s),\cdots, F(\omega + L \omega_s)]^{\mathrm{T}} \in \mathbb{C}^{2L+1}$. The sparsity of the spectrum $F(\omega)$ ensures that the vector $\mathbf{x}(\omega)$ is sparse for each $\omega \in [-\omega_s/2, \omega_s/2]$ provided that $\omega_s \geq 2\pi B$ (See Fig.~\ref{fig:mwc}(d)). Algorithms and conditions for recovering $\mathbf{x}(\omega)$ from $\mathbf{y}(\omega)$ are derived in \cite{mishali_2009, mishali_2010} where the authors showed that the minimum number of the hardware channels should be greater than the number of bands, that is, $M\geq 2N$. Hence, for reconstruction, the minimum sampling rate is $2NB$, which is twice the rate for unique identifiability. In practice, the number of channels can be reduced below $2N$, as low as one channel, by designing a specific set of periodic signals $p_m(t)$ \cite{mishali_2009}. 

\begin{figure}
\centering
\subfigure[\label{fig:LP}]
{\includegraphics[width=2.5in]{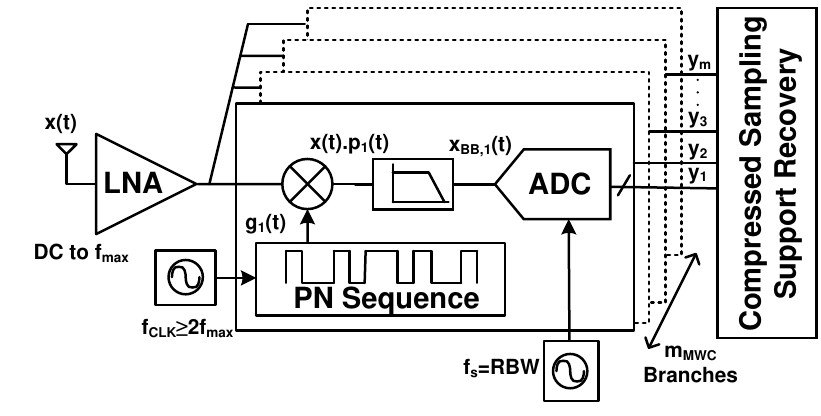}}
\hfill
\subfigure[\label{fig:BP}]
{\includegraphics[width=3in]{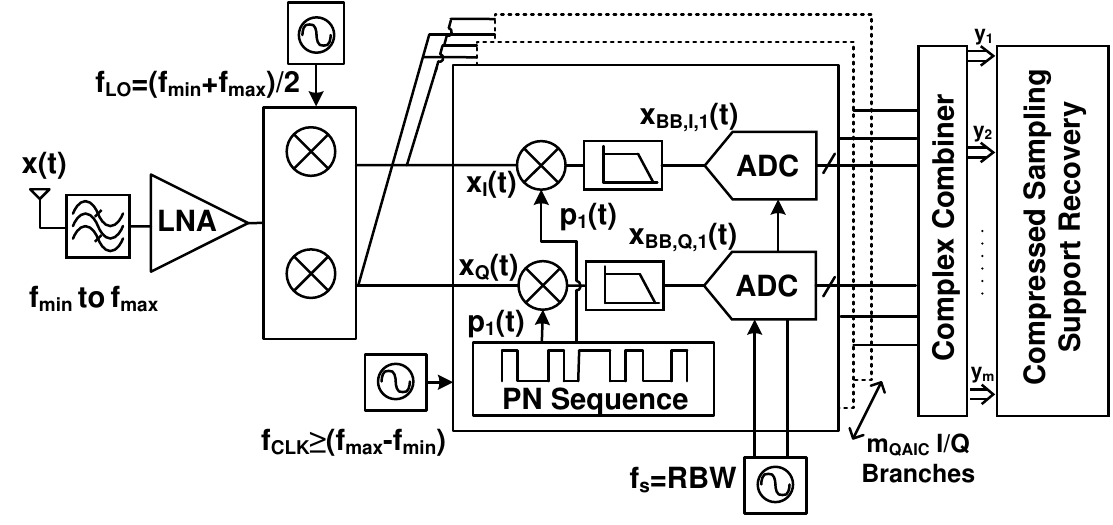}}
\hfill
\caption{Hardware prototypes for Sub-Nyquist sampling known as compressed sampling analog-to-information converter architectures; (a) Low-pass Modulated Wideband Converter (MWC)~\cite{mishali2011xampling, mishali2011bSPMag}, (b) Band-pass Quadrature Analog-to-Information Converter (QAIC)~\cite{Haque2015, JSSCYaz2015}.}
\label{fig:AIC}
\end{figure}

{\color{black}The periodic signals in an MWC are typically realized using a sign-alternating sequence or pseudorandom noise (PN) sequence generator. The PN sequence has to alternate signs a sufficiently large number of times within a time period to ensure that the sparse spectrum is estimated from the compressed measurements. Thus, these PN mixers operate at the Nyquist rate of the maximum input signal frequency. For example, for the multiband signal shown in Fig.~\ref{fig:multiband}, the mixers operate at $2\pi/T_{\text{Nyq}}$. By co-designing the analog preprocessing with digital signal processing, QAIC applies a band-pass filter and downconverts the high-frequency input signal. Specifically, by assuming that the spectra of the multiband signal lie in a frequency interval $[\omega_{\min}, \omega_{\max}]$, where $\omega_{\max} \leq 2\pi/T_{\text{Nyq}}$ and $\omega_{\min}>0$, the downconverted signal spectrum lie in the range $[0, \omega_{\max}-\omega_{\min}]$. With this change in the spectrum, the Nyquist rate of the signal is reduced by $\omega_{\min}$, and the same amount reduces the rate of the PN sequence. A comparison of the lowpass filter-based MWC and a bandpass filter-based QAIC is shown in Fig.~\ref{fig:AIC}. If $\omega_{\min} \gg 0$ for the application of interest, the power savings introduced by the band-pass QAIC architecture due to the reduction of the clock rate and sequence length of the PN sequence generator compared to the low-pass MWC architecture would be significant up to an order of magnitude for the analog preprocessing block interfacing with the sub-Nyquist sampling ADC as demonstrated in~\cite{JSSCYaz2015}. }

In the FRI and multiband frameworks, power reduction is achieved by reducing the sampling rates. However, in both these approaches, analog front ends are used to facilitate sub-Nyquist sampling. For example, in Fig.~\ref{fig:mwc}, $M$ channels are used where each channel consists of a multiplier, a lowpass filter, and a PN sequence generator. These circuits can be built by using existing low-power circuits and hence do not add significant power gain to the overall circuits.

\begin{figure}
    \centering
    \includegraphics[width = 2.5 in]{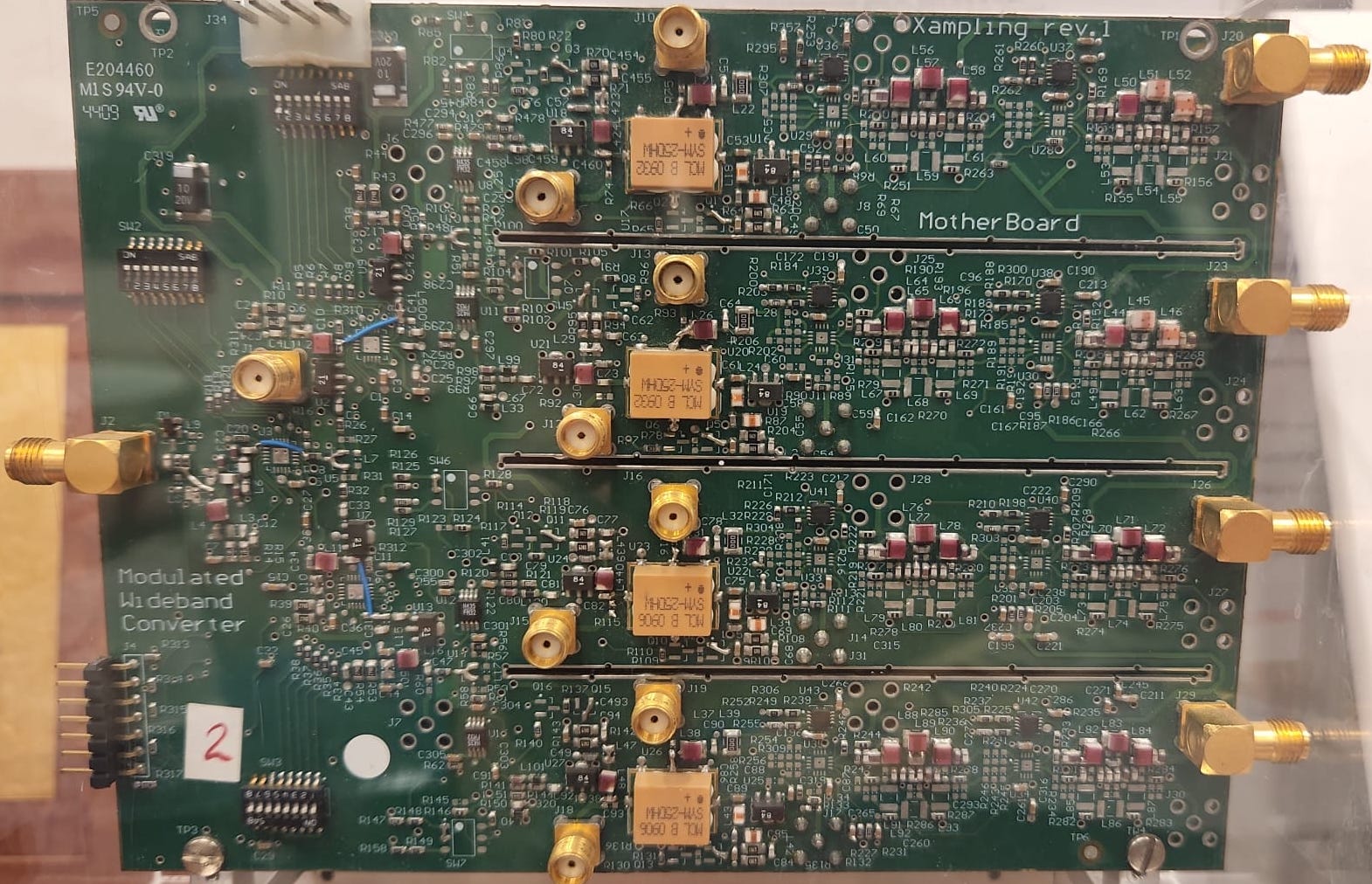}
    \caption{A hardware realization of the
MWC consisting of $M=4$ channels \cite{mishali2011xampling, mishali2011bSPMag}.}
    \label{fig:mwc_board}
\end{figure}

As in the case of FRI sampling, power saving is also applicable for multiband signals when the minimum sampling rate, $2NB$ Hz, is less than the Nyquist rate. Again, the savings amount depends on the actual sampling rate ratio to the Nyquist rate. In \cite{mishali2011xampling, mishali2011bSPMag}, Mishali et al. proposed hardware prototypes for MWC-based sub-Nyquist samplers for $B = 19$ MHz, $N = 6$, and $1/T_{\text{Nyq}} = 2$ GHz, the signals can be sampled and reconstructed at a rate of 280 MHz by using a $M=4$ channel MWC (See Fig.~\ref{fig:mwc_board}). By using the multiband structure, there is a seven-fold reduction in the sampling rate, which reduces the power requirements of the ADC by the same amount.

\section{Modulo-ADC}
\label{sec:modulo_adc}
The higher the DR of an ADC, the higher the number of bits required to keep a low quantization error, which results in high power consumption. Hence, it is desirable to keep the DR low. However, to avoid clipping due to saturation, the signal's DR must be within that of ADC's. In many applications, it is not always possible to premeditate the signal's DR and then choose an ADC with a suitable DR. Moreover, the available ADCs have a fixed DR in many scenarios. An attenuator can be applied to address these mismatches in DRs that scale down the signal to fit into ADC's DR. This solution may not be suitable when the amplitude of the signal varies significantly over time. For such a signal, small components of the signals are scaled down below the noise floor and hence cannot be detected.

\begin{figure}
    \centering
    \includegraphics[width = 2.5in]{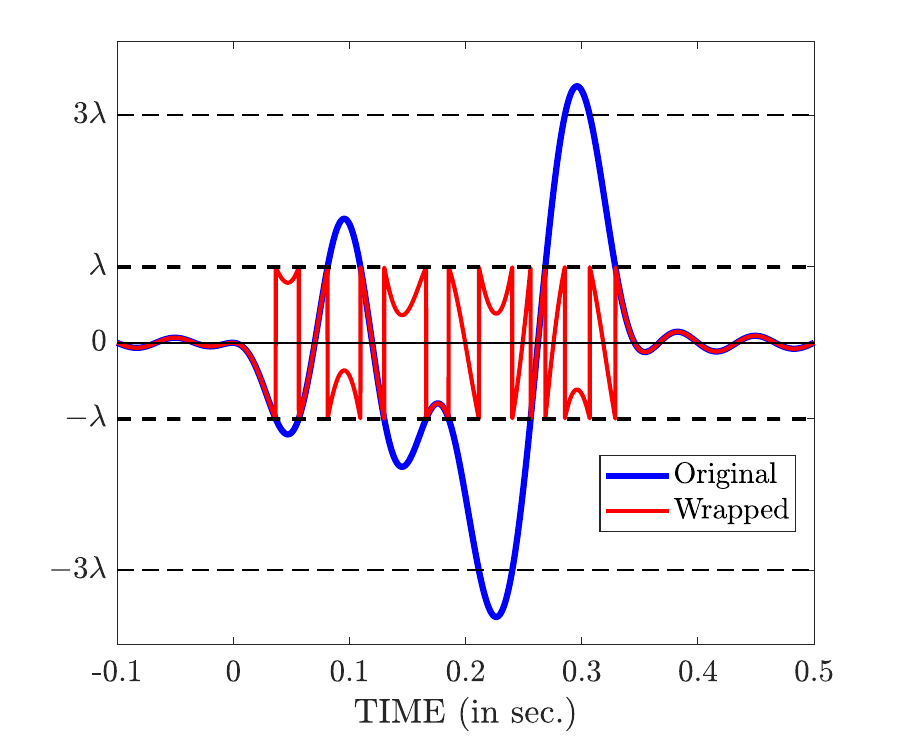}
    \caption{An illustration of modulo folding of a bandlimited signal; the ADC's DR is $[-\lambda, \lambda]$, and signal is folded to stay within the range. }
    \label{fig:modulo_folding}
\end{figure}

\begin{tcolorbox}[float*=t,
    width=2\linewidth,
	toprule = 0mm,
	bottomrule = 0mm,
	leftrule = 0mm,
	rightrule = 0mm,
	arc = 0mm,
	colframe = myblue,
	colback = mypurple,
	fonttitle = \sffamily\bfseries\large,
	title = Beyond Bandwidth Residue Recovery ($B^2R^2$) Algorithm ]	
    We give a brief overview of the $B^2R^2$ algorithm \cite{eyar_tsp} that can be used to unfold bandlimited signal samples. In the discussion, we assume that signals have finite energy and bandlimited to frequency interval $[-\omega_{\text{Nyq}}, \omega_{\text{Nyq}}]$. In addition, we assume that the sampling rate is above the Nyquist rate, that is, $\omega_s > 2\omega_{\text{Nyq}}$. The following two properties of a finite-energy bandlimited signal are used in this algorithm.
    \begin{enumerate}
        \item \emph{Time-domain decay property}: From the Riemann-Lebesgue Lemma we have that $\lim_{|t| \to \infty}f(t) = 0$. The time-domain decay implies that for every $\lambda >0$ there exists an integer $N_{\lambda}$ such that $|f(nT_s)| < \lambda,$ for all $|n|> N_{\lambda}$. 
            \item \emph{Frequency-domain decay property}: Since the signal is sampled above the Nyquist rate, the discrete-time Fourier transform of the samples $f(nT_s)$ are zero beyond the bandwidth, Mathematically, $F(e^{\mathrm{j}\omega T_s}) = 0$ for $\quad  \omega_{\text{Nyq}} < |\omega| < \omega_s / 2$.
         \end{enumerate}
    By using the decomposition in \eqref{eq:mod_decompos} and the time-domain decay property, we have that $z(nT_s) = 0, |n|> N_{\lambda}$ as $f_\lambda(nT_s) = f(nT_s)$ for those samples. 
    
    Further, by using the bandlimitedness of the samples, we get $F_{\lambda}(e^{\mathrm{j}\omega T_s}) = Z(e^{\mathrm{j}\omega T_s}), \quad  \omega \in \rho = (-\omega_s / 2, -\omega_{\text{Nyq}}) \cup (\omega_{\text{Nyq}}, \omega_s / 2).$ Hence, the samples of the residue are compactly supported over $n \in [-N_{\lambda}, N_{\lambda}]$, and its DTFT is known over a finite-length interval $\rho$. Let vectors $\mathbf{f}_\lambda, \mathbf{z} \in \mathbb{R}^{2N_{\lambda}+1}$ denote samples $f_{\lambda}(nT_s)$ and $z(nT_s)$, respectively, for $n \leq |N_{\lambda}|$. In addition, let $\mathcal{F}_{\rho}$ be a partial DTFT operator that evaluates Fourier measurements at frequencies in the set $\rho$. Then, from the above discussion, we can write the following optimization problem
      \begin{align}
            \underset{\mathbf{z}}{\min} \quad \mathrm{C}(\mathbf{z}) = \dfrac{1}{2}\|\mathcal{F}_{\rho}\mathbf{f}_{\lambda} - \mathcal{F}_{\rho}\mathbf{z}\|^2 \quad \text{s.t.} \quad \mathbf{z} \in \mathcal{S}_{N_\lambda},
        \label{eq:opt_b2r2}
        \end{align}
        where $\mathcal{S}_{N_\lambda}$ is a set of sequences that have support on the interval $[-N_{\lambda}, N_{\lambda}]$.

    Problem \eqref{eq:opt_b2r2} can be solved using a projected gradient descent (PGD) method starting from an initial point $\mathbf{z}^0 \in \mathcal{S}_{N_\lambda}$. The steps at the $k$-th iteration are
        \begin{equation}
            \begin{split}
                \mathbf{y}^{k} &= \mathbf{z}^{k-1} - \gamma_k \nabla \mathrm{C}(\mathbf{z}^{k-1}) \\
                \mathbf{z}^k &= P_{\mathcal{S}_{N_{\lambda}}}(\mathbf{y}^k),
            \end{split}
         \label{eq:pgd}
        \end{equation}
        where $\gamma_k >0$ is a suitable step-size, $\nabla \mathrm{C}(\mathbf{z}) = \mathcal{F}^*_{\rho}\mathcal{F}_{\rho}{(\mathbf{z} - \mathbf{f}_{\lambda})}$ is the gradient of $\mathrm{C(\mathbf{z})}$ and $P_{S_{N_\lambda}}(\cdot)$ is the orthogonal projection onto $\mathcal{S}_{N_{\lambda}}$. 

        After estimating $z(nT_s)$ from the modulo samples, the unfolded samples are determined by \eqref{eq:mod_decompos}.
        
\label{Box:b2r2}
\end{tcolorbox}	
 
\subsection{Theory and Algorithms}
A modulo-folding operation was suggested by Bhandari et al. \cite{uls_tsp} that folds the signal before sampling such that its variations are contained within the ADC's DR. To be precise, let $f(t)$ be an analog signal to be sampled, which is bounded as $|f(t)|\leq A$. The ADC's DR is $[-\lambda, \lambda]$ where $\lambda<A$. To process $f(t)$ through the ADC, it is first folded as $f_\lambda(t) = \mathcal{M}_\lambda \left(f(t) \right)$ where $\mathcal{M}(\cdot)$ is the folding operation defined as $\mathcal{M}_{\lambda}(a) = (a+\lambda)\,\, \text{mod}\,\, 2\lambda -\lambda$ for a real-valued $a$. The folding operation ensures that $|f_\lambda(t)| \leq \lambda$. The folded signal is then sampled using a low-DR ADC, which results in measurements $f_\lambda(nT_s)$ where $1/T_s$ is samples/sec. The folding operation does not discard any information of the analog signals; hence, the samples $f_\lambda(nT_s)$ retain necessary information of $f(t)$. In fact, mathematically, it was shown that bandlimited signals are uniquely identifiable from modulo samples provided that they are sampled above their Nyquist rate \cite{uls_tsp, eyar_tsp}. Based on this discussion, the key message is that the dynamic range of a modulo-ADC, an ADC preceded by a modulo operation, is much higher (theoretically unlimited) than the ADC's DR. Hence, the modulo-based approach is termed a high-DR ADC. An example of modulo folding for a bandlimited signal is shown in Fig.~\ref{fig:modulo_folding} where the signal with high-dynamic range is folded to stay within the ADC's DR.

In the preceding paragraph, we mentioned that modulo-DR's ADC is unlimited ($A/\lambda\rightarrow \infty$), provided that the sampling rate is higher than the Nyquist rate. However, in practice, recovery of $f(t)$ from the folded samples $f_\lambda(nT_s)$ depends on $A/\lambda$ and $T_s$. Typically, the reconstruction process works in two stages: first, determine the true samples $f(nT_s)$ from the folded samples by using an unfolding algorithm, and second, use standard reconstruction to get $f(t)$ from $f(nT_s)$. For bandlimited signals, the latter part requires that the sampling rate should be above the Nyquist rate. However, for unfolding, the signals are required to be oversampled where the oversampling factor $(\text{OF}=f_s/f_{\text{Nyq}})$ varies from algorithm to algorithm. Here $f_{\text{Nyq}}$ is the Nyquist rate.

Unfolding algorithms \cite{uls_tsp, bhandari2021unlimited, eyar_tsp, mulleti2022modulo} typically use the following decomposition of the modulo signal or samples.
\begin{align}
    f_\lambda(nT_s) = f(nT_s) + z(nT_s), \label{eq:mod_decompos}
\end{align}
where $z(t)$ is a piecewise constant function whose values are  multiple of $2\lambda$ and hence $z(nT_s) \in 2\lambda \mathbb{Z}$. Bhandari et al. \cite{uls_tsp} proposed a higher-order differences approach for unfolding. Let $\Delta^d$ be the $d$-th order sample difference operator. Then we have that $\mathcal{M}_\lambda(\Delta^d f_\lambda(nT_s)) = \mathcal{M}_\lambda \left(\mathcal{M}_\lambda(\Delta^d f(nT_s)) +  \mathcal{M}_\lambda(\Delta^d z(nT_s)) \right)$. The second term is zero as  $\Delta^d z(nT_s) \in 2\lambda \mathbb{Z}$. Hence, if 
\begin{align}
    |\Delta^d f_\lambda(nT_s) | < \lambda, \label{eq:diff_cond}
\end{align}
then we get $\mathcal{M}_\lambda(\Delta^d f_\lambda(nT_s)) = \Delta^d f(nT_s)$. In other words, we can determine the $d$-th order difference of the true samples from the folded samples provided that \eqref{eq:diff_cond} is satisfied. The authors in \cite{uls_tsp} showed that the condition is satisfied for $\text{OF} \geq 17$, and they proposed an approach to recover $f(nT_s)$ from higher-order differences. The sampling rate is high for this algorithm, and the use of sample differences makes it unstable in the presence of noise.

A robust and sample-efficient algorithm is proposed in \cite{eyar_tsp} where the authors used bandlimitedness and decay properties of the samples $f(nT_s)$ to extract them from $f_\lambda(nT_s)$. The main idea of the method and the corresponding optimization problem is discussed in the box \emph{Beyond Bandwidth Residue Recovery ($B^2R^2$) Algorithm}. In this algorithm, for an acceptable accuracy of the reconstruction, the OF varies between $2-8$ depending on the ratio $\theta = A/\lambda$ and the noise level. The modulo-ADC framework can also be extended to FRI and other classes of signals \cite{mulleti2022modulo, bhandari2022back}. For example, for the FRI sampling framework, a modulo folding operation is applied after the filtering stage (cf. Fig.~\ref{fig:fri_framework}), and an unfolding algorithm is used before the DSP block. As in the bandlimited case, it is required to sample above the rate of innovation for FRI signals when using a modulo-ADC. As a consequence, FRI sampling with modulo-ADC falls under the sub-Nyquist framework.

\begin{figure*}
    \centering
    \includegraphics[width = 7 in]{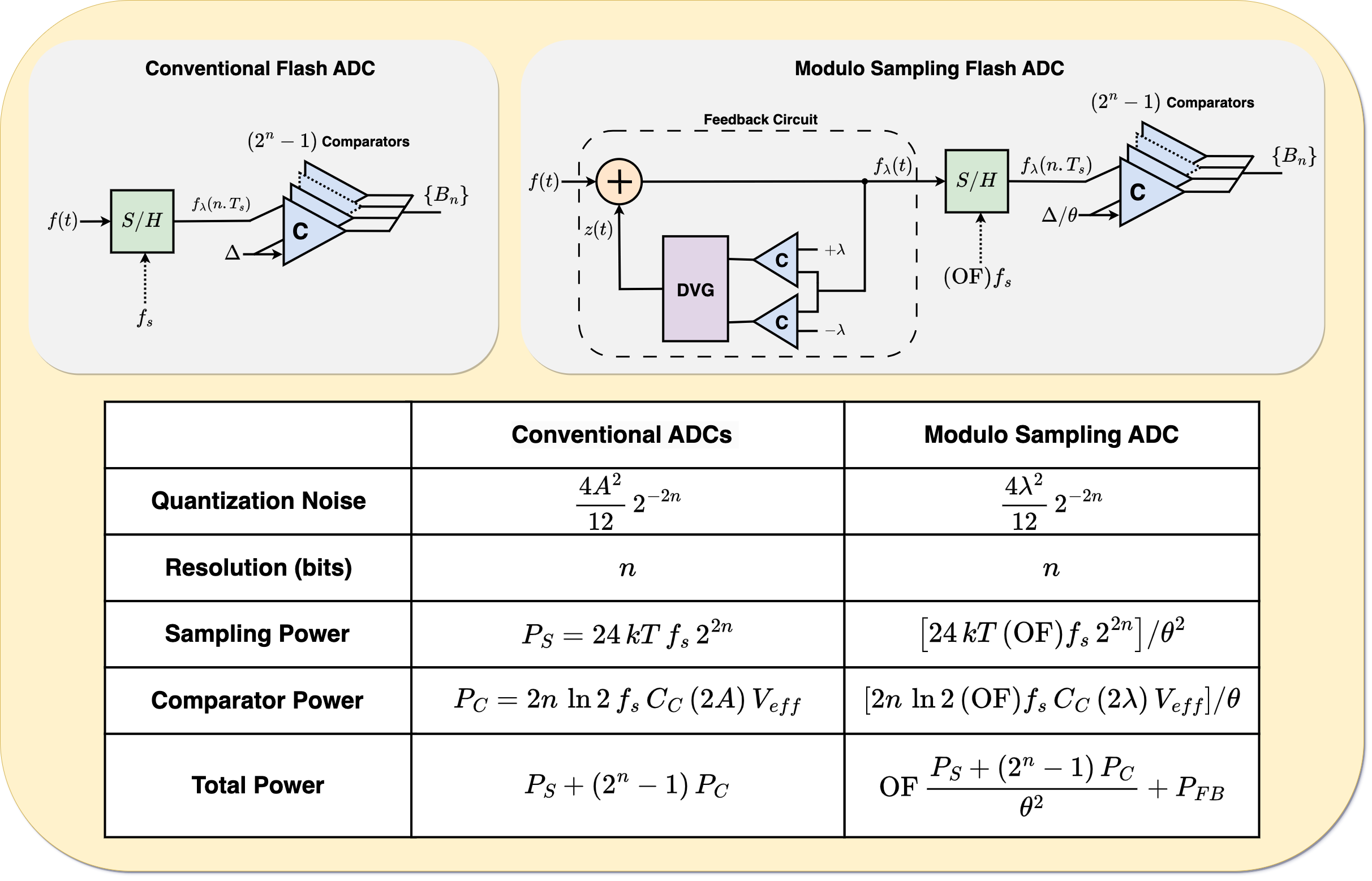}
    \caption{Power consumption in a modulo sampling ADC and conventional ADC.}
    \label{fig:modulo_hw}
\end{figure*}

\subsection{Power-Analysis of Modulo-ADC}
We derive theoretical lower bounds on the power consumption of a modulo-ADC and compare it with a conventional ADC. To this end, we consider a conventional ADC with sampling rate $f_s$ (which is equal to the Nyquist rate for a bandlimited signal or equal to RoI for an FRI signal) and input signals dynamic range $A$, as shown in Fig.~\ref{fig:modulo_hw}. For a $n$-bit flash-type quantizer, the total power consumption is given by \eqref{eq:adc_power}, which is also indicated in the table in Fig.~\ref{fig:modulo_hw}. To determine the power of the modulo-ADC, we determine the power consumption of its constituents: a modulo-folding circuit and a conventional low-DR ADC (cf. Fig.~\ref{fig:modulo_hw}). The low-DR ADC has a DR range $[-\lambda, \lambda]$ with $n$ bit resolution and operates at an oversampling rate $\text{OF}\, f_s$. Following the derivations presented in Section~\ref{sec:traditionalADC} and assuming that the capacitance values of the low-DR ADC are the same as that of the conventional ADC, the power of the sampler and the comparator of the low-DR ADC are given as 
\begin{align}
    \frac{24 kT\, \text{OF}\, f_s\,2^{2n}}{\theta^2}, \quad \text{and} \quad \frac{4n \ln 2 \, \text{OF}\, f_s C_C\, 2\lambda V_{eff}}{\theta}, \nonumber
\end{align}
respectively. By comparing these expressions with those of the conventional ADC, we note that the power consumption of the low-DR ADC $P^{'}$ is given as
\begin{align}
    P^{\prime} = \frac{\text{OF}}{\theta^2} P,
    \label{eq:mod_adc_part}
\end{align}
where $P = P_S + P_C$ is the power of a conventional ADC.

From the analysis up to this point, we infer that for $\text{OF}< \theta^2$, the S/H with quantization part of a modulo-ADC consumes less power than a conventional ADC. For example, in the modulo-ADC hardware prototype presented in \cite{mulleti2023hardware}, $\theta \leq 8$ and $\text{OF}\leq 5$, which is a factor of 12.8 reduction in power consumption. However, the feedback circuitry of a modulo-ADC will add additional power requirements, which we will consider next.

The goal of the modulo-circuit, or the feedback circuit, is to fold an input signal $f(t) \in [-A, A]$ to a dynamic range $[-\lambda, \lambda]$. The circuit consists of two comparators and a digital voltage generator (DVG). The input to the comparators is the modulo signal $f_\lambda(t)$. The comparators trigger whenever their input signal crosses $\pm \lambda$. The DVG reacts to the trigger signals and generates a residual signal $z(t) \in 2\lambda \mathbb{Z}$, which is added to the input signal to keep it within $[-\lambda, \lambda]$. Hence, we focus on the power utilization of the capacitors and the DVG. To this end, it is required to find an expression of the operating frequency $f_{FB}$ of the feedback loop. The frequency $f_{FB}$ depends on how fast the capacitors have to switch, and it could be higher than the sampling rate of the ADC ($=\text{OF} f_s$). Specifically, for any signal $f(t)$, $f_{FB}$ denotes an upper bound on the number of times it crosses amplitude levels ${\pm(2m+1), m \in \mathbb{N}}$. To derive an upper bound in $f_{FB}$ for a bandlimited signal, we use the fact that the number of zeros of a bandlimited function is equal to the Nyquist rate. Then, by using a derivation similar to that in \cite[Appendix]{mulleti2022modulo}, it can be shown that $f_{FB} = 2 \theta f_{\text{Nyq}}$ where $f_{\text{Nyq}}$ is the Nyquist rate. Based on $f_{FB}$, the power consumption of the comparators can be modeled similarly to \eqref{eq:flash_power}. However, the power consumption of the comparator is proportional to $\ln{\theta}$ instead of $2n\,\ln{2}$ since the comparators need to resolve the voltage of $2\lambda$ and generate a voltage level of $A$ at the output. Therefore, we rewrite the lower bound as \eqref{eq:comp2} where $C_{C,FB}$ is the latch comparator load capacitance. The power consumption of the DVG is modeled as the power consumption of a Capacitive Digital-to-Analog Converter (C-DAC) since it scales down or up the output voltage in multiples of $2\lambda$ with reference voltage $A$. We estimate the average power consumption of the DVG as \eqref{eq:DAC} where $C_{DAC}$ and $N$ are the total C-DAC capacitance and bit resolution of the C-DAC in the feedback circuit respectively \cite{5711005},

\begin{align}
        P_{\textit{C,FB}} =  \ln{\theta} \, f_{FB} \, C_{C,FB} \,  V_{\text{eff}} \, A,
        \label{eq:comp2}
\end{align}

\begin{align}
        P_{\textit{DVG}} = P_{\textit{DAC}} = 0.66\,\frac{f_{FB} \, C_{DAC} \, A^2}{N+1},
	\label{eq:DAC}
\end{align}

\begin{align}
        P_{\textit{FB}} = P_{\textit{DVG}} + 2 \, P_{\textit{C,FB}}. 
        \label{eq:fb_pwr}
\end{align}

The total power consumption of the feedback circuit shown in \eqref{eq:fb_pwr} is the sum of DVG and two comparators. The total power consumption of the modulo-ADC is given as 
\begin{align}
    P_{\text{mod-ADC}} = P^{\prime} + P_{\textit{FB}} = \frac{\text{OF}}{\theta^2} P + P_{\textit{FB}}.
\end{align}
For a fixed $f_s$, $A$, and $n$, we note that power reduction in the sampling and quantization part of the modulo-ADC is inversely proportional to $\theta^2$ whereas that in the modulo-circuit is proportional to $\theta \, \ln \theta$. The capacitance of the C-DAC can be minimized to reduce the power consumption of the DVG of a modulo-circuit under the noise and linearity limitations. For example, with a unit capacitance value of $10\,\thinspace\mathrm{f\/F}$, a 4-bit C-DAC can be used that allows up to $\theta = 16$. In this case, we have $C_{DAC}=2^4(10\,\thinspace\mathrm{f\/F})=160\,\thinspace\mathrm{f\/F}$. Now, if $A$ and $f_{FB}$ are assumed to be in $3.3\,\text{V}$ and $10\,\text{MHz}$ ranges, respectively, this results in $P_{\textit{DVG}} \simeq 2.3\thinspace\mathrm{\mu\/W}$ which is much less compared to the power consumption of the state-of-the-art ADC in \cite{adc_survey} that operates close to this frequency range. Hence, by carefully optimizing the feedback circuit, a modulo-ADC can have lower power requirements compared to a conventional ADC for any $\theta>1$. 

In the context of power saving for modulo-ADC, we would also like to emphasize that the oversampling factor, which depends on the unfolding algorithm, may not be independent of $\theta$. In fact, as shown via simulations in \cite{eyar_tsp}, the $\text{OF}$ for most algorithms increases with $\theta$. In such cases, the power saving in a modulo ADC still holds if $\text{OF}$ does not increase faster than $\theta^2$. Hence, designing sample-efficient algorithms plays a key role in the power-saving aspect of a modulo-ADC.

\begin{figure}
    \centering
    \includegraphics[width = 3 in]{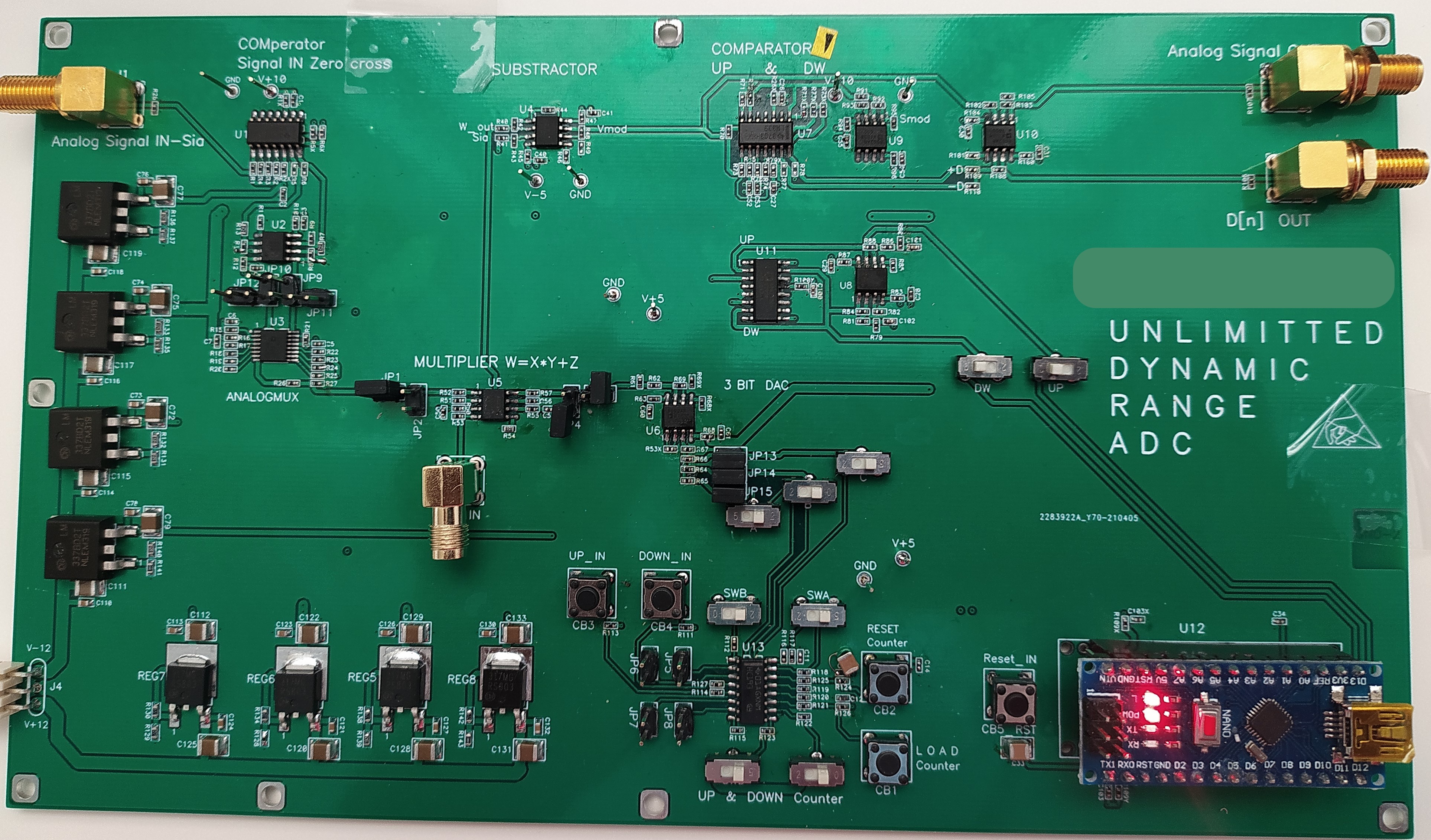}
    \caption{A hardware prototype of modulo-ADC \cite{mulleti2023hardware}.}
    \label{fig:mod_hw}
\end{figure}

A few works presented hardware prototypes of modulo-ADCs \cite{bhandari2021unlimited,bhandari2022back, mulleti2023hardware}. The emphasis in these works is to show that one can fold the analog signal to a low-dynamic range and the applicability of the unfolding algorithms. In \cite{bhandari2021unlimited,bhandari2022back}, the authors consider different signal models, hardware limitations, and algorithms; however, they did not provide details of the hardware implementation. Moreover, the results are presented for signals up to $300$ Hz. A hardware prototype, shown in Fig.~\ref{fig:mod_hw}, for signals up to 10 kHz, with detailed circuitry, is presented in \cite{mulleti2023hardware}. The authors showed that, for FRI signals, the system operates six times below the Nyquist rate, and the modulo ADC can sample signals that are eight times larger than the ADC's DR.

\section{Asynchronous Sampling}
\label{sec:asynchronous_sampling}
Sampling approaches discussed in the previous sections are based on uniform sampling where a continuous-time signal $f(t)$ is represented by $f(nT_s), T_s>0, n \in \mathbb{Z}$. A key advantage of uniform sampling is that it allows the application of standard Fourier analysis. This, in turn, relates the spectra of $f(t)$ and $f(nT_s)$ and often leads to closed-form expressions for reconstruction, such as Shannon-Nyquist recovery. Despite being widely used, a shortcoming of uniform sampling is that it measures samples at the same rate, usually dictated by the maximum signal component, and ignores whether the signal is changing rapidly or not locally. Specifically, the instantaneous bandwidth of most real-world signals varies across time, even if their overall bandwidth is fixed. In such scenarios, it is power-efficient to measure samples at a low rate in the regions with small changes and vice-versa.

\begin{figure}
    \centering
    \includegraphics[width= 2.8in]{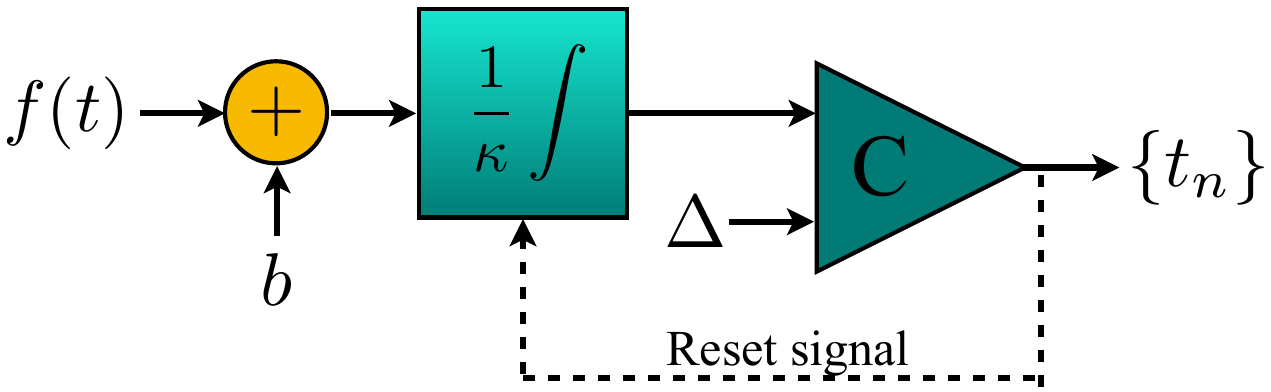}
    \caption{A schematic of IF-TEM}
    \label{fig:iftem}
\end{figure}

To this end, event-based samplers such as
level-crossing sampling 
\cite{logan1977information}
and integrated and fire time-encoding machines (IF-TEM) \cite{lazar2004perfect,gontier2014sampling,alexandru2019reconstructing,hila_tsp} are applied where the sampling locations are a function of the analog signal. In these frameworks, analog signals are represented by a set of time instants $\{t_n\}$ instead of their values at a specific time instants as in uniform and random sampling. For example, in a popular level-crossing sampling architecture called zero-crossing detector, the signal is represented by its zero-crossing instants \cite{logan1977information}. In general level-crossing, a signal is represented by time instants at which the signal crosses a set of predefined amplitude levels.

\begin{figure}
    \centering
    \includegraphics[width = 2 in]{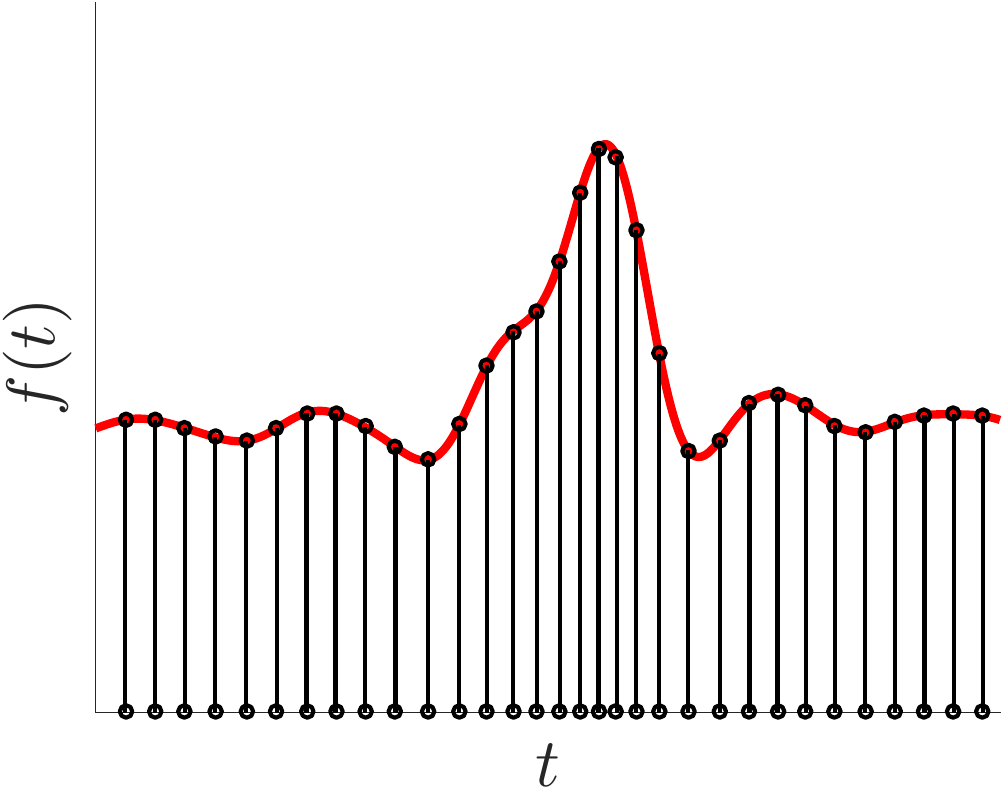}
    \caption{IF-TEM samples of a bandlimited signal: the samples are dense when the signal changes rapidly compared to the regions of slow variations.}
    \label{fig:iftem_samples}
\end{figure}
\begin{tcolorbox}[float*=t,
    width=2\linewidth,
	toprule = 0mm,
	bottomrule = 0mm,
	leftrule = 0mm,
	rightrule = 0mm,
	arc = 0mm,
	colframe = myblue,
	colback = mypurple,
	fonttitle = \sffamily\bfseries\large,
	title = FRI Signal Reconstruction From IF-TEM Samples ]	
    To determine FRI parameters from IF-TEM samples, we assume that the FRI signal is filtered through an SoS kernel as in \eqref{eq:sos} and the output signal $y(t)$ (cf. \eqref{eq:sos_op}) is sampled using IF-TEM. Let the time encodings be $\{t_n\}_{n=1}^N$. Using \eqref{eq:iftem1}, we compute the following amplitude measurements from time encodings: $y_n = \kappa \Delta - b(t_n - t_{n-1}) = \int_{t= t_{n-1}}^{t_n} y(t) \, \mathrm{d}t, \quad n = 1, \cdots, N-1$. From trigonometric polynomial form of $y(t)$ in \eqref{eq:sos_op}, the Fourier samples are related to the measurements as
    \begin{align}
        y_n = \sum_{k \in \{-K, \cdots, K\} \setminus 0} F(k\omega_0) \frac{\left( e^{\mathrm{j}k\omega_0 t_{n+1}}- e^{\mathrm{j} k\omega_0 t_{n}} \right)}{\mathrm{j} k\omega_0}+F(0) \left( t_{n+1} - t_{n}\right), \quad n = 1, \cdots, N-1, \label{eq:yn1}
    \end{align}
    where $K\geq L$.
     It was shown that the Fourier samples $\{F(k\omega_0)\}_{k=K}^K$ are uniquely determined from the measurements $\{y_n\}_{n = 1}^{N-1}$ using \eqref{eq:yn1} if $N \geq 2K+2$ \cite{hila_tsp}. The condition $N \geq 2K+2$ can be satisfied by choosing the IF-TEM parameters as $\text{FR}_{\min} > \frac{2K+1}{t_{\max}}$ where $t_{\max}$ is maximum time delays of the FRI signal. The latter condition is similar to that for reconstructing bandlimited signals from IF-TEM samples where it is required that $\text{FR}_{\min}$ is greater than the Nyquist rate. 

        After the Fourier samples are obtained from the time encodings, the FRI parameters can be computed using either Prony's approach or other robust methods, as discussed previously.

\label{Box:fri_tem}
\end{tcolorbox}	
 
\section{Theory and Algorithms}
The level-crossing-based approach requires multiple comparators, one for each level, which could be inefficient for hardware implementation, although more power-efficient implementations were demonstrated \cite{fixed_w_jssc13}. On the other hand, IF-TEMs use a single comparator to time encode analog signals. In an IF-TEM, an analog signal $f(t)$ is added to a bias $b$ such that $b+f(t)\geq 0$. This is ensured by choosing the bias as $b>A$ where $|f(t)|\leq A$. Thus, bias $b$ is determined by the input signal dynamic range. The positive signal is scaled as $\frac{1}{\kappa}(b+f(t))$ where $\kappa>0$ and then integrated. The resulting signal is compared with a threshold $\Delta$ as shown in Fig. \ref{fig:iftem}. When the value of the output of the integrator reaches threshold $\Delta$, defined by input signal level, bias $b$, and signal bandwidth, the time instant is recorded, and the integral is reset. The process is repeated to get the time encodings $\{t_n\}$, which are digital representations of $f(t)$ and is a function of IF-TEM parameters $\{b, \kappa, \Delta\}$. As discussed next, the reconstruction of $f(t)$ is based on non-uniform sampling-based recovery methods.

Following the sampling mechanism of IF-TEM discussed above, $f(t)$ and its time encodings are related as
\begin{align}
	\frac{1}{\kappa} \int_{t= t_{n-1}}^{t_n}\left(b+f(t)\right)\, \mathrm{d}t = \Delta.
	\label{eq:iftem1}
\end{align}
From this equation and by using the boundedness $|f(t)|<c$, the difference between consecutive firings are related as
\begin{align}
	\frac{\kappa \Delta }{(b+A)} \leq t_n -t_{n-1} \leq \frac{\kappa \Delta }{(b-A)}.
	\label{eq:firing_rate}
\end{align}
By using these bounds, the maximum and minimum number of time encodings per second or \emph{firing rates} (FR) are given as
\begin{align}
	\text{FR}_{\max} = \frac{(b+A)}{\kappa \Delta }, \quad \text{and} \quad \text{FR}_{\min} = \frac{(b-A)}{\kappa \Delta },
	\label{eq:firing_rate1}
\end{align}
respectively. The notion of FR has a similar meaning to that of sampling rate in uniform sampling or sampling density (average number of samples per second) in random sampling. The difference is that FR is signal-dependent, whereas sampling density or rate is fixed. The FR is higher when the signal changes rapidly compared to when there are slow variations in the signal. An illustrative example is shown in Fig.~\ref{fig:iftem_samples}.

For perfect reconstruction of $f(t)$, the IF-TEM parameters should be chosen such that $\text{FR}_{\min}$ is above a certain rate. For example, Lazar and T\'oth \cite{lazar2004perfect} used an iterative algorithm for the reconstruction of bandlimited signals and showed that the algorithm converges to the true signal provided that $\text{FR}_{\min}$ is above the Nyquist rate. Similarly, \cite{hila_tsp} showed that FRI signals could be reconstructed from their time encodings if the minimum firing rate is set above the rate of innovation. The reconstruction approach is summarized in the box \emph{FRI Signal Reconstruction From IF-TEM Samples. }

\begin{figure*}
    \centering
    \includegraphics[width = 5 in]{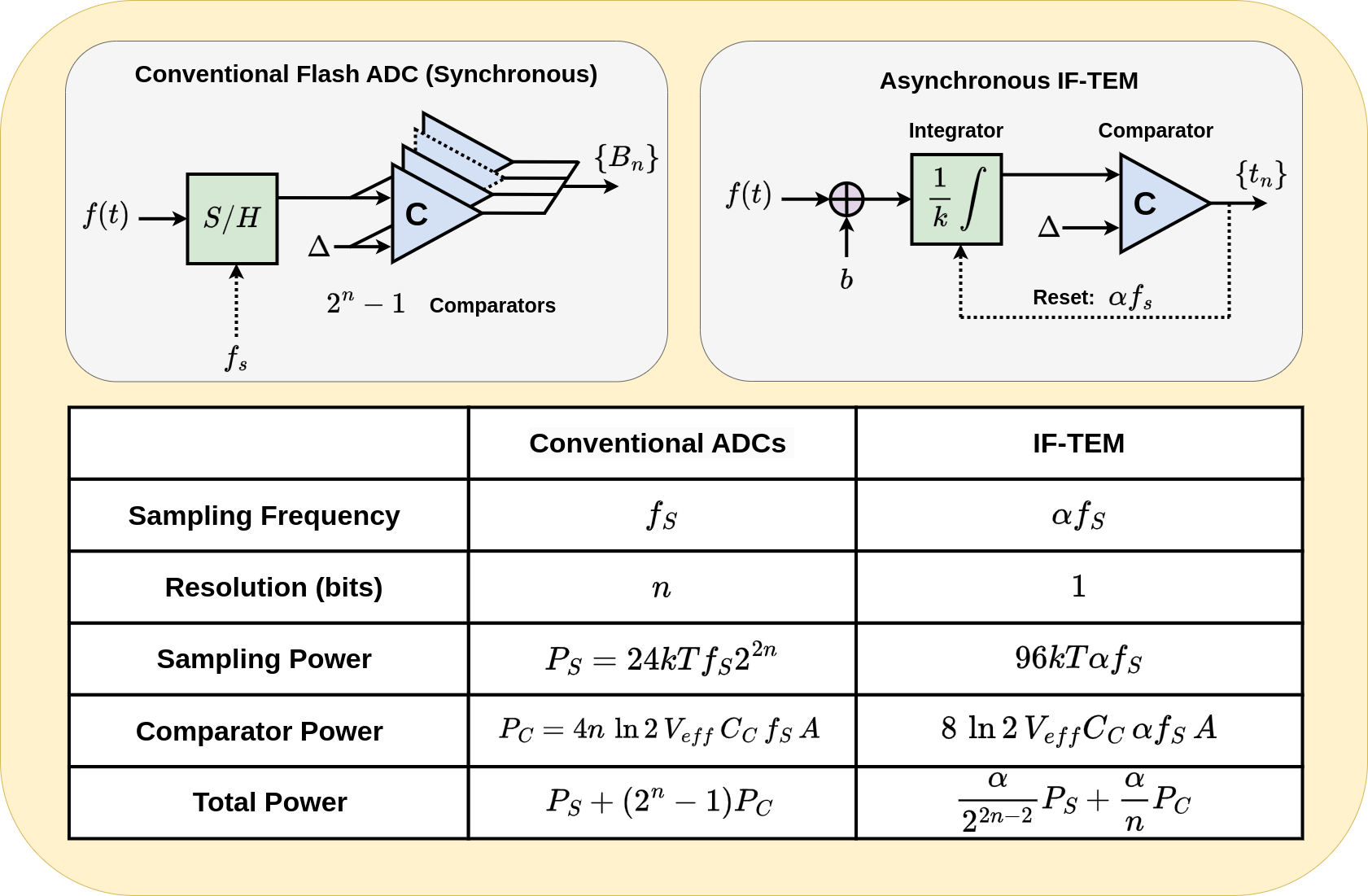}
    \caption{Architecture and power consumption estimation comparison between conventional Flash ADCs and IF-TEM, showing corresponding power savings for the latter system.}
    \label{fig:iftem_table}
\end{figure*}
\begin{figure}
    \centering
    \includegraphics[width = 3in]{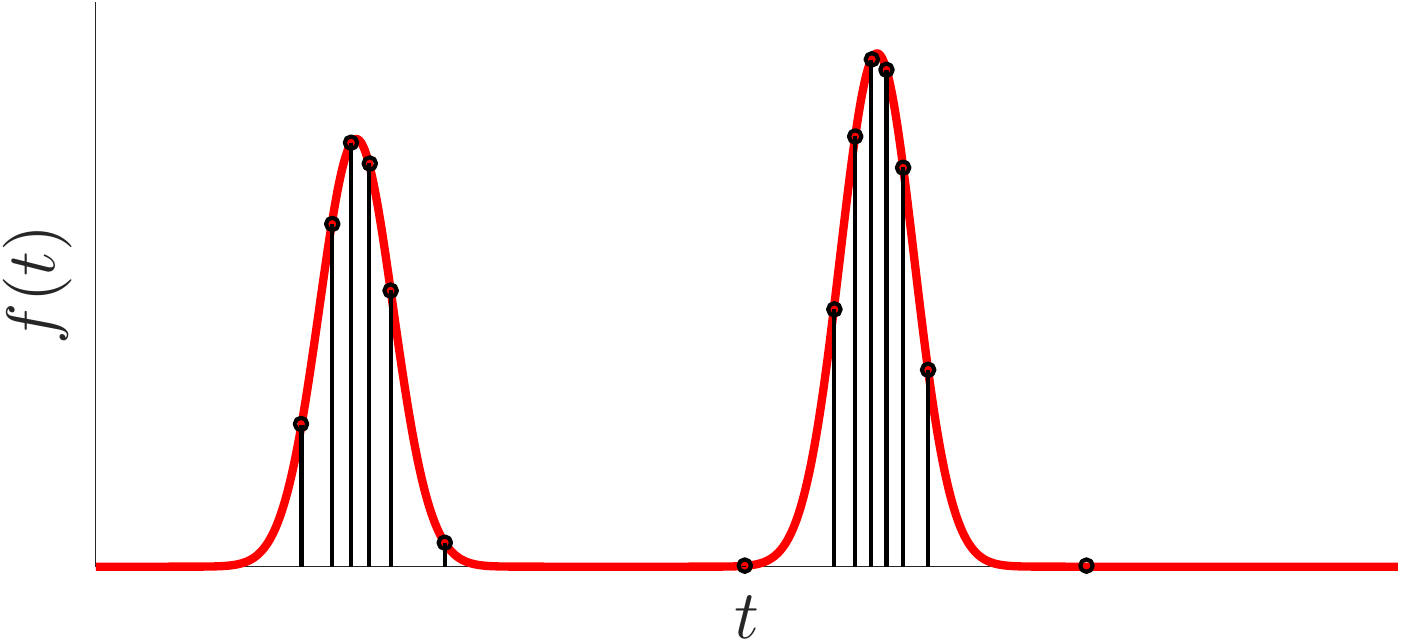}
    \caption{IF-TEM sampling by using burst signals with positive amplitudes: We observe that there are no samples when the signal is inactive, whereas conventional ADC would sample during the entire time interval. }
    \label{fig:iftem_samples1}
\end{figure}

\subsection{Power Efficiency of IF-TEMs}
In the following, we determine the power an IF-TEM circuit consumes and compare it with that in a conventional ADC for the same input signal. The details of our analysis are depicted in Fig.~\ref{fig:iftem_table}. For the conventional ADC, we use \eqref{eq:adc_power} to represent its power consumption. For the IF-TEM, we note that there are three components: (i) an adder with a bias, (ii) an integrator with a reset rate $\alpha f_s$, where $\alpha$ is a non-uniform sampling activity factor that reflects non-periodical sampling activity,  and (iii) a comparator with a threshold $\Delta$. Among these, the integrator and comparator are power-hungry blocks, and we shall focus on them in the following analysis.

Quantization of time-encodings utilized in IF-TEMs is usually not carried out in the same way as in conventional ADCs, but specialized time-to-digital circuits (TDCs) are used \cite{szyduczynski2023time}. The underlying principle in TDCs is that time can be divided into measurable intervals that are quantized into digital representations or bits. There are several ways of realizing a TDC, such as analog-based, counter-based, or delay-line. (please refer to \cite{szyduczynski2023time} for details). Among these, most digital implementation-based TDCs have an order of magnitude lower power consumption than typical analog circuits. This is the reason we did not account for the TDC in the power calculation of the IF-TEM framework.

For the calculation of the power consumption, we note that the integrator can be viewed as an active sample-and-hold (a capacitor and a switch) with $n=1$ bit resolution \cite{vandePlassche2003}, since the number of comparators defines the system resolution, similar to conventional flash ADCs. Hence, integrator power consumption is given as
\begin{align}
    P_{\text{INT}} = 96 kT \alpha f_s,
    \label{eq:p_int}
\end{align}
where $\alpha f_s$ is the switching or reset rate of the integrator. It is purposely written in terms of the sampling rate $f_s$ of the conventional ADC (See Fig.~\ref{fig:iftem_table}). The sampling activity factor $\alpha$ shows how the non-uniform switching rate of an integrator relates to the sampling frequency of conventional ADCs. For example, in the case when the average switching rate of an integrator approaches the conventional ADC sampling frequency  $f_s$ in case of a fast-changing signal, as illustrated in  Fig.~\ref{fig:iftem_samples}, $\alpha$ approaches or goes above the value of 1. If the average sampling rate (switching rate) of an IF-TEM is much lower than that of the conventional ADC in the case of burst signals, with activity only in some time intervals (as in Fig.~\ref{fig:iftem_samples1}), then $\alpha \ll 1$, an important case for event-driven low-power data conversion systems. Our power analysis reflects these changes in the activity factor.

Overall, The sampling activity factor $\alpha >0$ is a quantity that depends on the signal to be sampled and the reconstruction methods, as discussed in the following. The reset rate has a similar notion as the $f_{\text{FB}}$ in the case of modulo-ADC discussed in Section~\ref{sec:modulo_adc}. To compute the comparator's power consumption, we use \eqref{eq:flash_power} with an assumption that the comparator has the same characteristics as the comparators in the flash ADC. Then, for $n=1$, the comparator power is given as
\begin{align}
    P_{C, \text{IF-TEM}} = 8 \, \ln2 \,V_{\text{eff}} \,C_C \, \alpha f_s  A. 
\end{align}
The total estimated power of the IF-TEM is
\begin{align}
    P_{\text{IF-TEM}} &= P_{\text{INT}} + P_{C, \text{IF-TEM}}, \nonumber \\
     & = 96 kT \alpha f_s + 8 \, \ln2 \,V_{\text{eff}}\, C_C \, \alpha f_s  A, \nonumber \\
     & = \frac{\alpha}{2^{2n-2}} P_{S} + \frac{2\alpha}{(2^n-1)n}  \left((2^n-1) P_C \right), \label{eq:power_iftem}
     \end{align}
where $P_S$ and $(2^n-1)P_C$ are power requirements of the S/H and quantizer parts (cf. \eqref{eq:sh_power}), respectively, in a conventional ADC. We assumed that the comparators were the same for both the conventional ADC and the IF-TEM.  

From \eqref{eq:power_iftem}, we note that for $\alpha < \max( 2^{2n-2}, (2^{n}-1)n)$, where $n>1$, $P_{IF-TEM}$ is smaller than the power of a conventional ADC. The power saving is largely due to the fact that there is only a single comparator in the IF-TEM compared to the conventional flash-ADC, where there are $2^n-1$. Even if one uses a different quantizer than the flash-type one, the number of comparators can be reduced at the expense of other circuitry; still, the power saving of IF-TEM can be reduced if $\alpha <1$. Hence, in the aforementioned analysis, $\alpha$ plays a key role, and next, we discuss its dependency on the class of signals and reconstruction algorithm. 

First, we consider bandlimited signals whose Nyquist rate is $f_{\text{Nyq}}$, and hence the conventional ADC operates at the minimum rate $f_s = f_{\text{Nyq}}$. For these signals, as discussed earlier, reconstruction from IF-TEM samples is possible if $\text{FR}_{\min} > f_{\text{Nyq}}$ \cite{lazar2004perfect}. If for a given signal's dynamic range $A$, we choose IF-TEM parameters $\{b, \kappa,\Delta\}$ such that $\text{FR}_{\min} > f_{\text{Nyq}}$ is satisfied, then from \eqref{eq:firing_rate1} we know that the firing rate could be much higher than the Nyquist rate. Specifically, in $\alpha > \frac{b+A}{b-A} >1$. This implies that IF-TEM always fires even if the input signal's amplitude does not change significantly. This fact is also depicted in Fig.~\ref{fig:iftem_samples}. For bandlimited signals, a large value of $\alpha$ above one reduces the effective power saving even if $\alpha < \max( 2^{2n-2}, (2^{n}-1)n)$.

\begin{figure}
    \centering
    \includegraphics[width = 2.5 in]{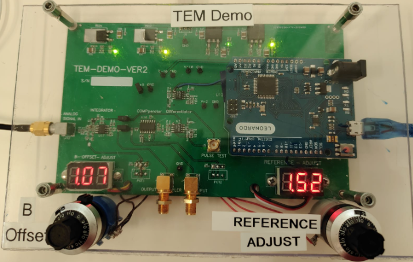}
    \caption{A hardware prototype of IF-TEM Sampler \cite{naaman2023hardware}.}
    \label{fig:tem_hw}
\end{figure}

On the other hand, signals that are not necessarily bandlimited but are bursty in nature, where the signal's amplitude is above a certain level only in a few intervals, are much more suited for sampling by using an IF-TEM. Assume that the reconstruction is not a sinc-based interpolation as in the Shannon-Nyquist framework but uses a local interpolation such as a linear or a spline-based one. In such scenarios, $\text{FR}_{\min}$ can be set to zero, which implies there are no firings when the signal is not active, and the signal is reconstructed, up to a given accuracy, from the time encodings in the active period. In this case, $\alpha < 1$. An example of such a burst signal, which is similar to an FRI signal, is shown in Fig.~\ref{fig:iftem_samples1}, where the signal is active in certain time intervals and zero elsewhere. We assume that the signal is positive such that $0\leq f(t)\leq A$ and hence bias is not added, that is, $b=0$. Then, for $\kappa=1$ and a particular value of $\Delta$, we note that there are no firings during the inactive period of the signal. This implies that the integrator and the comparator were not activated, i.e., in sleep mode, in those intervals. However, if a conventional ADC is used to sample such signals, then they operate in both active and inactive modes and consume relatively higher power. Because of this reason, time-encoding-based ADCs, including IF-TEMs, are preferable in event detection applications such as bio-signal monitoring in wearable devices, voice-activity detection by voice-based virtual assistant platforms, and more. On another application front, even-based cameras are an emerging field where it is shown that they have higher resolution and lower power consumption than conventional cameras.

Several integrated circuit implementations of level-crossing-based systems are available \cite{wake-up_jssc21}. However, there is very little work on IF-TEMs. In \cite{naaman2023hardware}, a hardware prototype of an IF-TEM ADC was presented, where reconstruction of FRI signals from time encodings was demonstrated (see Fig.~\ref{fig:tem_hw}). The authors showed that the firing rate is ten times below the Nyquist rate of the signals.

\section{ Low-Bit Quantization}
\label{sec:low-bit quantization}
\begin{figure}[t]
    \centering
    \includegraphics[width = 3 in]{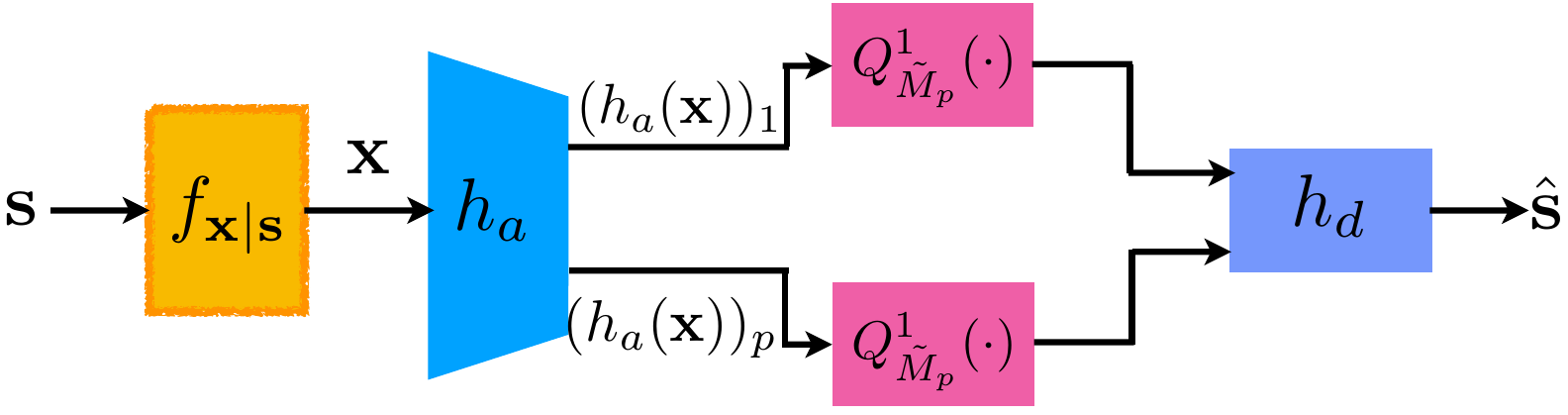}
    \caption{A general framework of task-based quantizer}
    \label{fig:task_based}
\end{figure}

 In the previous sampling frameworks, the focus was on reducing the sampling rate and the dynamic range of the signal to reduce power consumption. In this section, we discuss methods to reduce the number of quantization bits by exploring the task at hand. 
For signal reconstruction with acceptable accuracy, samples should be encoded with a sufficiently large number of bits. However, in many applications, the task is not signal reconstruction but extracting some information from the analog signal \cite{shlezinger2019hardware}. For example, in a MIMO communication system, it is necessary to estimate the channel, which is a low-dimensional task compared to underlying analog signal recovery. In such applications, task-based quantization was proposed where by using analog combiners and joint optimization of an analog front end and digital back end, the number of quantization bits can be significantly reduced \cite{shlezinger2019hardware}.

\subsection{Theoretical Framework of Task-Based Quantizer}
To understand the task-based low-bit quantization system, let us consider a discrete model where we assume that the signals are already sampled \cite{shlezinger2019hardware}. The assumption helps us in focusing only on the quantization part. The general framework of task-based quantization is shown in Fig.~\ref{fig:task_based} where the goal is to estimate a low-dimensional vector or task $\mathbf{s}\in \mathbb{R}^K$ from measurements $\mathbf{x}\in \mathbb{R}^N$ where $K<N$. The information $\mathbf{s}$ is statistically dependent on observations $\mathbf{x}$ via a conditional distribution $f_{\mathbf{x}|\mathbf{s}}$. For example, $\mathbf{x}$ represents directions of arrivals, and $\mathbf{y}$ is a snapshot of $N$ sensors or antennas for the case of radio receivers. Instead of individually quantizing entries of $\mathbf{x}$ by using $N$ scalar quantizers, the measurement $\mathbf{y}$ is projected to a low-dimensional space by using a combiner $h_a: \mathbb{R}^N \rightarrow \mathbb{R}^P$. Each component of the combiner, $(h_a(\mathbf{x}))_p, 1\leq p \leq P$ is independently quantized by a scalar quantizer $Q_{M_P}^1(\cdot)$ to get quantized measurements $Q_{M_P}^1\left((h_a(\mathbf{x}))_p\right), 1\leq p \leq P$. Here, $M_P$ represents the number of levels of each quantizer. The vector $\mathbf{s}$ is estimated from the quantized measurements as 
\begin{align}   
	\hat{\mathbf{s}}=h_d\left(	Q_{\tilde{M}_p}^1\left((h_a(\mathbf{x}))_1\right),\cdots,Q_{\tilde{M}_p}^1\left((h_a(\mathbf{x}))_p\right)
	\right),
	\label{Eq:task_systems_s_hat}
\end{align}
where $h_d(\cdot):\mathcal{R}^p \longmapsto \mathcal{R}^k$ is an estimator. The aforementioned system is hardware-limited by assuming that the number of quantization levels available, $M$, is fixed. This implies that, $M_P$ should satisfy the inequality $M_P\leq M^{\frac{1}{P}}$.

The goal of a hardware-limited task-based system is to jointly optimize the combiner $h_a(\cdot)$, quantizer $Q_{M_P}$, and estimator $h_d(\cdot)$. The problem is intractable without making explicit assumptions about the stochastic model $f_{\mathbf{x}|\mathbf{s}}$. For example, one can consider a linear model $\mathbf{x} = \mathbf{A}\mathbf{s} + \mathbf{w}$ where $\mathbf{w}$ is noise \cite{shlezinger2019hardware}. Such models are common in many estimation tasks. Further, the combiner and estimator are assumed to be linear and represented by matrices. Within this linear setting, authors in \cite{shlezinger2019hardware} derived expressions for the combiner matrix and estimator that minimizes mean-squared error in the estimation of $\mathbf{s}$ for a fixed $M$. The results are also extended to the measurements related to $\mathbf{s}$ in a quadratic fashion \cite{salamatian_2019}. In a nutshell, by reducing the dimension of the measurements from $N$ to $P$, one can assign more bits during quantization when the overall number of bits is fixed. This improves the estimation accuracy.

Deep learning-based methods are also applied when the observations and unknowns, such as channel parameters, can not be represented linearly or quadratically \cite{Shlezinger_Entropy21}. Further, at the application front, task-based methods are applied to massive MIMO communications for estimation of the underlying channel from the high-dimensional received signals \cite{Shlezinger_Tsp19Asym}.


\begin{tcolorbox}[float*=b,
    width=2\linewidth,
	toprule = 0mm,
	bottomrule = 0mm,
	leftrule = 0mm,
	rightrule = 0mm,
	arc = 0mm,
	colframe = myblue,
	colback = mypurple,
	fonttitle = \sffamily\bfseries\large,
	title = Task-Specific MIMO Recovery Algorithm ]	

Task-specific MIMO receiver design constraints can be accounted for in the convex optimization framework by formulating a loss measure: 
\begin{equation}
\label{eqn:Loss1}
    \mySet{L}(\myMat{A}) = {\rm MSE}(\myMat{A}) +  \gamma_I {\rm IntRej}(\myMat{A}) + \gamma_S \|\myMat{A}\|_{1,1}.
\end{equation}
Here $\|\cdot\|_{1,1}$ is the entry-wise $\ell_1$ norm operator, while $\gamma_I, \gamma_S \geq 0$ are regularization coefficients, balancing the contribution of task recovery MSE, spatial interferer rejection, and sparsity level of the analog combiner in the overall loss measure. In addition, although \eqref{eqn:Loss1} is expressed in the convex term, the final solution, accounting for finite VM resolution, is defined over a discrete domain space. The final objective is formulated as: 

\begin{equation}
\label{eqn:OptProblem}
    \myMat{A}  = \underset{\myMat{A} \in \mySet{A}^{P \times N}}{\arg \min}{\mySet{L}(\myMat{A})}
\end{equation}

 
The recovery algorithm performs $k_{\max}$ rounds of a convex optimizer for minimizing $\mySet{L}(\myMat{A})$ over $\mySet{C}^{P\times N}$, with periodic projections onto the discrete $\mySet{A}$. Algorithm~\ref{alg:Algo1} summarizes the design procedure.

\label{Box:ts_mimo}
\end{tcolorbox}

\subsection{Task-Based ADC Hardware}
A hardware prototype for MIMO channel reduction with the task of estimating the underlying channel is presented in \cite{gong_nir_hw}.

One of the main goals of the low-bit quantization using a task-based (Task-Specific) framework is to reduce the overall system's power consumption. Implications of task-based receiver system design are considered in detail in this section. First of all, as previously mentioned, the power consumption of the ADCs is directly proportional to the number of bits, as suggested by Walden's FoM \cite{walden_adc}. 

The task-based framework discussed in the earlier section was shown to reduce the quantization rate quite significantly (x2-6 times). However, additional analog processing is required for such systems, resulting in increased analog hardware complexity directly tied to power consumption, which is especially critical in MIMO receivers since analog processing overhead is proportional to the hardware processing channels. In fact, power efficiency, hardware complexity, and signal recovery accuracy are critical performance trade-offs for MIMO modern communication systems. 


\begin{figure}
    \centering
    \includegraphics[width = 3 in]{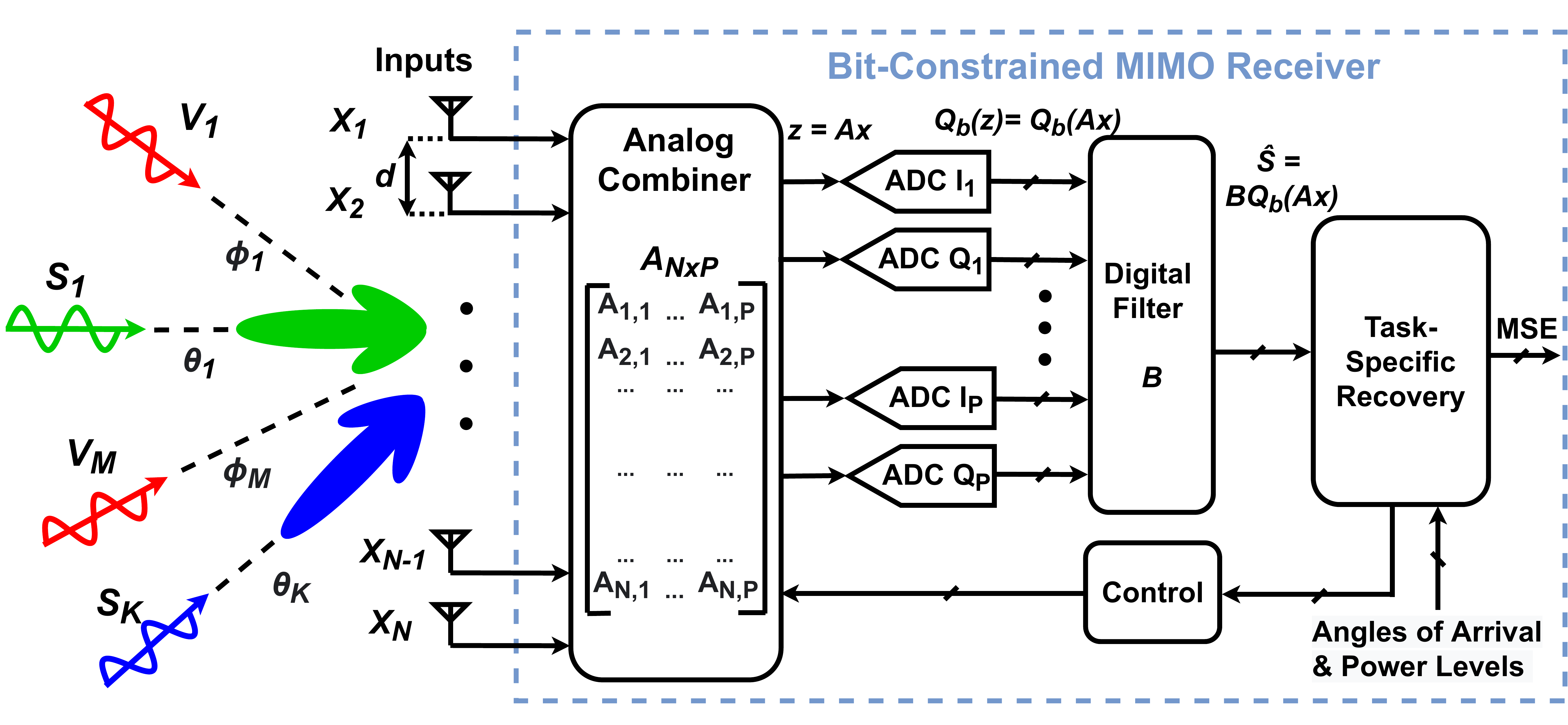}
    \caption{Task-Specific hybrid MIMO receiver system utilizing low-quantization rate ADCs and power-efficient analog combiner \cite{zirtiloglu_ts_mimo}.}
    \label{fig:ts_mimo}
\end{figure}

In \cite{zirtiloglu_ts_mimo}, energy-efficient analog processing together with low-quantization ADCs and task-specific processing in application to MIMO receivers, was developed. The recent MIMO developments particularly focus on communication in congested environments, where both desired signals and undesired jammers arrive from various directions at the receiver.

Generally, the signal model at the MIMO receiver can be described as 
vector $\myVec{x}$ is an input observation vector consisting of a linear combination of $K$ desired signals (\textit{tasks}) arriving from corresponding angles $\theta_K$. In addition, a set of $M$ undesired signals (spatial blockers) arriving from respective angles $\phi_K$, as shown in Fig. \ref{fig:ts_mimo}. Input observation signals are first processed in the analog combiner $\myMat{A}$, performing linear matrix multiplication with signal dimensionality reduction, similar to the $h_a(\cdot)$ combiner in the task-based quantization system discussed earlier (Fig. \ref{fig:task_based}). The output of the analog combiner is quantized using low-resolution ADCs (modeled as a scalar quantizer $Q_{b}(\cdot)$) and processed in the digital domain using matrix  $\myMat{B}$, similar to estimator $h_d(\cdot)$, obtaining desired signal (task) estimate $\hat{\myVec{s}} = \myMat{B} \mySet{Q}_b(\myMat{A}\myVec{x})$.

Several hardware design considerations were accounted for in the solution algorithm. First, the system targets low-bit ADC \textit{task} recovery without compromising recovery accuracy. Second, while MSE focuses on \textit{task} recovery, spatial blocker impression in the analog domain before analog-to-digital conversion is embedded to avoid receiver desensitization and increased dynamic range requirement of the ADCs. Finally, the system targets a highly power-efficient operation of the analog combiner to reduce the overhead from analog pre-processing. The details of the Task-Specific MIMO optimization algorithm are provided in the box and algorithm \emph{Task-Specific MIMO Recovery}.

\begin{algorithm}[!t]
	\caption{Analog combiner setting}
	\label{alg:Algo1}
	\KwData{Fix $\myMat{A}^{(0)}$ }
	\For{$k=1,2,\ldots,k_{\max}$}{
		Update $\myMat{A}^{(k)} \leftarrow \mySet{O}_{\mySet{L}}\big(\myMat{A}^{(k-1)}\big)$\\
		\If{$\mod(k,k_{\rm proj}) =0 $}{
		Project via $\myMat{A}^{(k)} \leftarrow \mySet{P}_{\mySet{A}}(\myMat{A}^{(k)})$
		}
	}
	\KwOut{Analog combiner $\myMat{A}^{(k_{\max})}$.}
\end{algorithm}

\begin{figure*}
    \centering
    \includegraphics[width = 7 in]{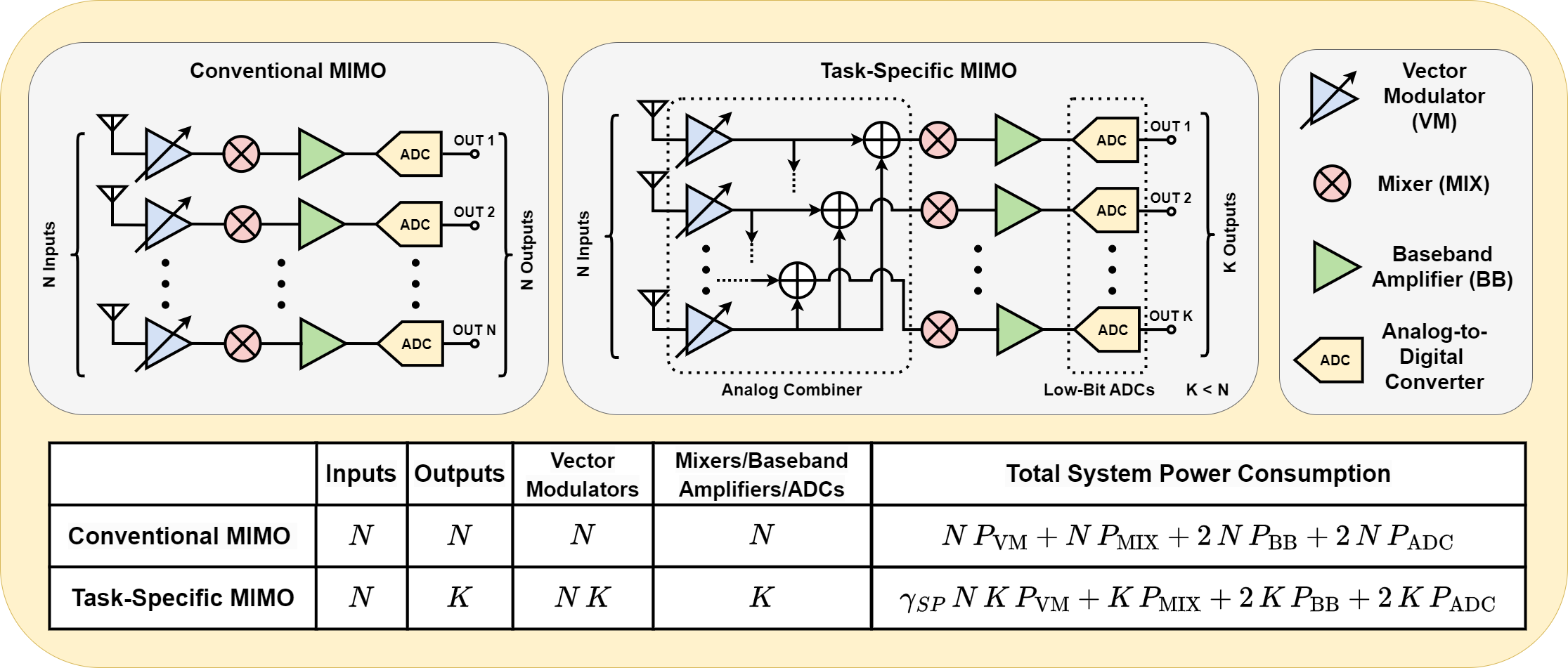}
    \caption{Architecture and power consumption estimation comparison between conventional MIMO and Task-Specific MIMO.}
    \label{fig:mimo_table}
\end{figure*}

\begin{figure}
    \centering
    \includegraphics[width = 3 in]{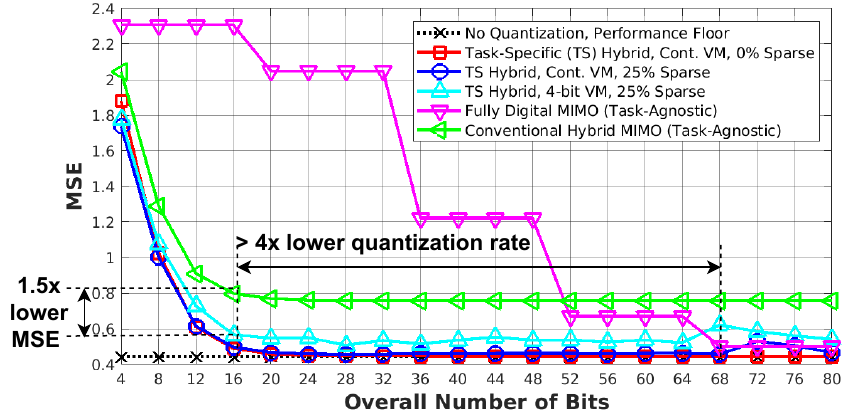}
    \caption{MSE vs. the total number of quantization bits for recovering 2 desired signals in the presence of 2 spatial interferers \cite{zirtiloglu_ts_mimo}. Task-specific solutions are compared to conventional hybrid and fully digital recovery approaches.}
    \label{fig:ts_mimo_sim}
\end{figure}

 The power consumption of a MIMO receiver can be estimated from the power consumption of the sub-blocks of the system. A conventional MIMO receiver hardware, shown in Fig. \ref{fig:mimo_table} (top left), consists of an analog signal acquisition block, such as low-noise amplifiers, variable gain amplifiers, and phase shifters that perform some form of signal processing in the analog domain, for example, for beamforming. For analysis simplicity, we considered an implementation using Vector Modulators (VMs), a circuit component that applies simultaneous complex gain and phase shift on the input signal. The next crucial component is a mixer, which performs signal downconversion from carrier to baseband frequencies. Signals are then amplified using baseband amplifiers and quantized using ADCs. With $N$ input signals and $N$ output streams, conventional MIMO system power consumption is expressed as: 
 \begin{equation}
    \label{eqn:power_fulldig}
      P_{\rm Conv\,MIMO} = N\, P_{\rm VM} + N \,
     P_{\rm MIX} + 2 \,N \, P_{\rm BB} + 2\,N\, 
    P_{\rm ADC},
\end{equation}
where $P_{\rm VM}$ is the power consumed by the vector modulator, $P_{\rm MIX}$ is the downconversion mixer consumption, and $P_{\rm BB}$ and $P_{\rm ADC}$ are baseband amplifiers and ADCs power consumption, respectively, each doubled to account for the quadrature I and Q paths.  

The Task-Specific MIMO receiver, on the other hand, first processes input signals in the analog combiner, which performs analog pre-processing with dimensionality reduction. For a system with $N$ input signals and $K$ output streams, where $K<N$, the number of required VMs is $N \times K$. However, due to the lower number of output signals, only $K$ baseband processing chains are needed. In addition to that, one has to consider optimal analog pre-processing circuit design, such that the power overhead from the pre-quantizing analog combiner would not exceed power savings enabled by the low quantization rate. Overall analog combiner power consumption depends on the number of active VM elements in $\myMat{A}$ and individual VM power consumption, which is proportional to the VM resolution. By utilizing sparsification of  $\myMat{A}$ (zeroing particular elements of matrix $\myMat{A}$), implemented as keeping the corresponding VM in sleep mode, and incorporating fine and coarsely quantized VMs, additional power savings in the analog combiner can be achieved. Hardware non-idealities, in this case, are accounted for in the back-end convex optimization algorithm.

The resulting system power consumption expression is: 
\begin{align}
    \label{eqn:power_ts_mimo}
    P_{\rm TS\,MIMO} = \gamma_{SP} \, N \, K & \, P_{\rm VM} + K \, P_{\rm MIX} \nonumber\\  &+ 2 K\, P_{\rm BB} + 2 K\, P_{\rm ADC},
\end{align} 
 Here, $\gamma_{SP}$ is the analog combiner sparsity coefficient: $\gamma_{SP} = 1$ denotes a non-sparse $\myA$, while, a practical value of $\gamma_{SP} = 0.75$, which corresponds to 25$\%$ sparsity, was used in the system. Fig. \ref{fig:mimo_table} features Task-Specific MIMO receiver diagram  (top right) and system parameters with power consumption estimation expressions. It is important to note that Task-Specific MIMO incorporates additional power savings from low-resolution VMs and ADCs. 

The proposed solution implementation was simulated in Matlab in \cite{zirtiloglu_ts_mimo}. The numerical results in Fig.~\ref{fig:ts_mimo_sim} show that the quantization rate can be reduced by a factor of 4 compared to conventional fully digital MIMO for the same accuracy. In addition, power consumption estimation based on state-of-the-art integrated components shows that more than 58$\%$ power reduction can be achieved compared to the task-agnostic solution by incorporating sparsity of the analog combiner, variable-resolution VMs, and low-resolution ADCs while suppressing interferers by more than 36 dB \cite{zirtiloglu_ts_mimo}.

Regarding the ADC power consumption estimation, we recall from the discussion above that typical Flash ADC power is estimated as  $P_{\textit{SH}} +  (2^n-1)P_{\textit{C ADC}},$ where $n$ is the number of bits. Although theoretical power-saving gain should be exponentially proportional to the reduction of the number of bits, practical ADC implementations exhibit power consumption overhead coming from other peripheral circuits (digital processing and biasing network), especially at a low resolution when consumption of those sub-blocks becomes more dominant. This creates a power consumption floor for low-resolution ADCs, as demonstrated in the ADC performance survey \cite{adc_survey}.

\section{Acknowledgement}
We would like to acknowledge several authors' publications and their research works that are relevant to this review. We could not cite them due to a strict restriction on the number of references.
\section{Conclusion}
\label{sec:outlook}
Low-power ADCs are key for emerging applications. In this review, we discussed four approaches that can be applied to reduce the power consumption of ADCs. Among these techniques, three directly reduce power by reducing the sampling rate, dynamic range, and resolution of an ADC, whereas the fourth method relies on a time-based discrete representation of the signal. Any two or more of these methods can be combined together to make the sampling process even more power efficient. A common feature of the four frameworks is to have a preprocessing analog front-end, a low-power ADC, and a digital signal processor. For all these four frameworks, we discussed power consumption and hardware prototypes that are developed in our labs. These power-efficient ADCs play a crucial role in designing compact, portable medical and communication devices.

\bibliographystyle{IEEEtran}
\bibliography{US_biblios,refs,refs2,tem_refs,tem_ref2,taskbased_refs,multiband, AIC}

\begin{thebibliography}{10}
\providecommand{\url}[1]{#1}
\csname url@samestyle\endcsname
\providecommand{\newblock}{\relax}
\providecommand{\bibinfo}[2]{#2}
\providecommand{\BIBentrySTDinterwordspacing}{\spaceskip=0pt\relax}
\providecommand{\BIBentryALTinterwordstretchfactor}{4}
\providecommand{\BIBentryALTinterwordspacing}{\spaceskip=\fontdimen2\font plus
\BIBentryALTinterwordstretchfactor\fontdimen3\font minus
  \fontdimen4\font\relax}
\providecommand{\BIBforeignlanguage}[2]{{%
\expandafter\ifx\csname l@#1\endcsname\relax
\typeout{** WARNING: IEEEtran.bst: No hyphenation pattern has been}%
\typeout{** loaded for the language `#1'. Using the pattern for}%
\typeout{** the default language instead.}%
\else
\language=\csname l@#1\endcsname
\fi
#2}}
\providecommand{\BIBdecl}{\relax}
\BIBdecl

\bibitem{adc_power}
T.~Sundstrom, B.~Murmann, and C.~Svensson, ``Power dissipation bounds for
  high-speed nyquist analog-to-digital converters,'' \emph{IEEE Transactions on
  Circuits and Systems I: Regular Papers}, vol.~56, no.~3, pp. 509--518, 2009.

\bibitem{walden_adc}
R.~H. {Walden}, ``Performance trends for analog to digital converters,''
  \emph{IEEE Comm. Mag.}, vol.~37, no.~2, pp. 96--101, Feb. 1999.

\bibitem{eldar_2015sampling}
Y.~C. Eldar, \emph{Sampling Theory: Beyond Bandlimited Systems}.\hskip 1em plus
  0.5em minus 0.4em\relax Cambridge University Press, 2015.

\bibitem{uls_tsp}
A.~{Bhandari}, F.~{Krahmer}, and R.~{Raskar}, ``On unlimited sampling and
  reconstruction,'' \emph{IEEE Trans. Signal Process.}, vol.~69, pp.
  3827--3839, 2020.

\bibitem{bhandari2021unlimited}
A.~Bhandari, F.~Krahmer, and T.~Poskitt, ``Unlimited sampling from theory to
  practice: {F}ourier-{P}rony recovery and prototype {ADC},'' \emph{IEEE Trans.
  Signal Process.}, vol.~70, pp. 1131--1141, 2022.

\bibitem{eyar_tsp}
E.~Azar, S.~Mulleti, and Y.~C. Eldar, ``Robust unlimited sampling beyond
  modulo,'' \emph{arXiv preprint arXiv:2206.14656}, 2022.

\bibitem{mulleti2022modulo}
S.~Mulleti and Y.~C. Eldar, ``Modulo sampling of {FRI} signals,'' \emph{arXiv
  preprint arXiv:2207.08774}, 2022.

\bibitem{logan1977information}
B.~F. Logan~Jr, ``Information in the zero crossings of bandpass signals,''
  \emph{Bell Sys. Technical J.}, vol.~56, no.~4, pp. 487--510, 1977.

\bibitem{lazar2004perfect}
A.~A. Lazar and L.~T. T{\'o}th, ``Perfect recovery and sensitivity analysis of
  time encoded bandlimited signals,'' \emph{IEEE Trans Circuits and Syst. I:
  Regular Papers}, vol.~51, no.~10, pp. 2060--2073, 2004.

\bibitem{gontier2014sampling}
D.~Gontier and M.~Vetterli, ``Sampling based on timing: Time encoding machines
  on shift-invariant subspaces,'' \emph{Applied and Comput. Harmonic Anal.},
  vol.~36, no.~1, pp. 63--78, 2014.

\bibitem{alexandru2019reconstructing}
R.~Alexandru and P.~L. Dragotti, ``Reconstructing classes of non-bandlimited
  signals from time encoded information,'' \emph{IEEE Trans. Signal Process.},
  vol.~68, pp. 747--763, 2019.

\bibitem{hila_tsp}
H.~Naaman, S.~Mulleti, and Y.~C. Eldar, ``{FRI-TEM}: {T}ime encoding sampling
  of finite-rate-of-innovation signals,'' \emph{IEEE Trans. Signal Process.},
  vol.~70, pp. 2267--2279, 2022.

\bibitem{shlezinger2019hardware}
N.~Shlezinger, Y.~C. Eldar, and M.~R. Rodrigues, ``Hardware-limited task-based
  quantization,'' \emph{IEEE Trans. Signal Process.}, vol.~67, no.~20, pp.
  5223--5238, Aug. 2019.

\bibitem{salamatian_2019}
S.~Salamatian, N.~Shlezinger, Y.~C. Eldar, and M.~M\'edard, ``Task-based
  quantization for recovering quadratic functions using principal inertia
  components,'' in \emph{2019 IEEE International Symposium on Information
  Theory (ISIT)}, 2019, pp. 390--394.

\bibitem{mishra_subnyquist_radar}
K.~V. Mishra, Y.~C. Eldar, E.~Shoshan, M.~Namer, and M.~Meltsin, ``A cognitive
  sub-{N}yquist {MIMO} radar prototype,'' \emph{IEEE Trans. Aerosp. Electron.
  Syst.}, vol.~56, no.~2, pp. 937--955, 2020.

\bibitem{mishali2011xampling}
M.~Mishali, Y.~C. Eldar, O.~Dounaevsky, and E.~Shoshan, ``Xampling: Analog to
  digital at sub-{N}yquist rates,'' \emph{IET Circ. Devices \& Sys.}, vol.~5,
  no.~1, pp. 8--20, 2011.

\bibitem{mishali2011bSPMag}
M.~Mishali and Y.~C. Eldar, ``Sub-{N}yquist sampling: {B}ridging theory and
  practice,'' \emph{IEEE Signal Process. Mag.}, vol.~28, no.~6, pp. 98--124,
  2011.

\bibitem{bhandari2022back}
A.~Bhandari, ``Back in the {US-SR}: {U}nlimited sampling and sparse
  super-resolution with its hardware validation,'' \emph{IEEE Signal Process.
  Lett.}, vol.~29, pp. 1047--1051, 2022.

\bibitem{mulleti2023hardware}
S.~Mulleti, E.~Reznitskiy, S.~Savariego, M.~Namer, N.~Glazer, and Y.~C. Eldar,
  ``A hardware prototype of wideband high-dynamic range {ADC},'' \emph{(to
  appear) IET Circuits Devices Syst.}, 2023.

\bibitem{naaman2023hardware}
H.~Naaman, N.~Glazer, M.~Namer, D.~Bilik, S.~Savariego, and Y.~C. Eldar,
  ``Hardware prototype of a time-encoding sub-{N}yquist {ADC},'' \emph{arXiv
  preprint arXiv:2301.02012}, 2023.

\bibitem{gong_nir_hw}
T.~Gong, N.~Shlezinger, S.~S. Ioushua, M.~Namer, Z.~Yang, and Y.~C. Eldar,
  ``{RF} chain reduction for {MIMO} systems: {A} hardware prototype,''
  \emph{IEEE Sys. J.}, vol.~14, no.~4, pp. 5296--5307, 2020.

\bibitem{yazicigilSPM}
R.~T. Yazicigil, T.~Haque, P.~R. Kinget, and J.~Wright, ``Taking compressive
  sensing to the hardware level: {B}reaking fundamental radio-frequency
  hardware performance tradeoffs,'' \emph{IEEE Signal Process. Mag.}, vol.~36,
  no.~2, pp. 81--100, 2019.

\bibitem{adc_survey}
B.~Murmann, ``{ADC Performance Survey 1997-2022},'' [Online]. Available:
  \url{https://github.com/bmurmann/ADC-survey}.

\bibitem{mishali2011SPMag}
M.~Mishali and Y.~C. Eldar, ``Wideband spectrum sensing at sub-{N}yquist
  rates,'' \emph{IEEE Signal Process. Mag.}, vol.~28, no.~4, pp. 102--135,
  2011.

\bibitem{eldar2009SPMag}
Y.~C. Eldar and M.~Mishali, ``Beyond bandlimited sampling,'' \emph{IEEE Signal
  Process. Mag.}, vol.~26, no.~3, pp. 48--68, 2009.

\bibitem{timur_power}
T.~Zirtiloglu, N.~Shlezinger, Y.~C. Eldar, and R.~Tugce~Yazicigil,
  ``Power-efficient hybrid {MIMO} receiver with task-specific beamforming using
  low-resolution {ADCs},'' in \emph{Proc. IEEE Int. Conf. Acoust., Speech and
  Signal Process. (ICASSP)}, 2022, pp. 5338--5342.

\bibitem{vetterli}
M.~Vetterli, P.~Marziliano, and T.~Blu, ``Sampling signals with finite rate of
  innovation,'' \emph{IEEE Trans. Signal Process.}, vol.~50, no.~6, pp.
  1417--1428, Jun. 2002.

\bibitem{fri_strang}
P.~L. Dragotti, M.~Vetterli, and T.~Blu, ``Sampling moments and reconstructing
  signals of finite rate of innovation: Shannon meets {S}trang-{F}ix,''
  \emph{IEEE Trans. Signal Process.}, vol.~55, no.~5, pp. 1741--1757, May 2007.

\bibitem{bluspmag}
T.~Blu, P.-L. Dragotti, M.~Vetterli, P.~Marziliano, and L.~Coulot, ``Sparse
  sampling of signal innovations,'' \emph{{IEEE} Signal Process. Mag.},
  vol.~25, no.~2, pp. 31--40, Mar. 2008.

\bibitem{eldar_sos}
R.~Tur, Y.~C. Eldar, and Z.~Friedman, ``Innovation rate sampling of pulse
  streams with application to ultrasound imaging,'' \emph{IEEE Trans. Signal
  Process.}, vol.~59, no.~4, pp. 1827--1842, Apr. 2011.

\bibitem{mulleti_kernal}
S.~{Mulleti} and C.~S. {Seelamantula}, ``Paley--{W}iener characterization of
  kernels for finite-rate-of-innovation sampling,'' \emph{IEEE Trans. Signal
  Process.}, vol.~65, no.~22, pp. 5860--5872, Nov. 2017.

\bibitem{lin_ppv_98}
Y.-P. Lin and P.~Vaidyanathan, ``Periodically nonuniform sampling of bandpass
  signals,'' \emph{IEEE Trans. Circuits Syst. {II}}, vol.~45, no.~3, pp.
  340--351, 1998.

\bibitem{herley_wong}
C.~Herley and P.~W. Wong, ``Minimum rate sampling and reconstruction of signals
  with arbitrary frequency support,'' \emph{IEEE Trans. Info. Theory}, vol.~45,
  no.~5, pp. 1555--1564, 1999.

\bibitem{mishali_2009}
M.~Mishali and Y.~C. Eldar, ``Blind multiband signal reconstruction: Compressed
  sensing for analog signals,'' \emph{IEEE Trans. Signal Process.}, vol.~57,
  no.~3, pp. 993--1009, Mar. 2009.

\bibitem{mishali_2010}
------, ``From theory to practice: Sub-{N}yquist sampling of sparse wideband
  analog signals,'' \emph{IEEE J. Sel. Topics in Signal Process.}, vol.~4,
  no.~2, pp. 375--391, Apr. 2010.

\bibitem{stoica}
P.~Stoica and R.~L. Moses, \emph{Introduction to Spectral Analysis}.\hskip 1em
  plus 0.5em minus 0.4em\relax Upper Saddle River, NJ: Prentice Hall, 1997.

\bibitem{blu_moms}
T.~Blu, P.~Thevenaz, and M.~Unser, ``{MOMS}: {M}aximal-order interpolation of
  minimal support,'' \emph{IEEE Trans. Image Process.}, vol.~10, no.~7, pp.
  1069--1080, Jul. 2001.

\bibitem{landau}
H.~J. Landau, ``\BIBforeignlanguage{English}{Necessary density conditions for
  sampling and interpolation of certain entire functions},''
  \emph{\BIBforeignlanguage{English}{Acta Mathematica}}, vol. 117, no.~1, pp.
  37--52, 1967.

\bibitem{JSSCYaz2015}
R.~T. Yazicigil and et~al., ``Wideband rapid interferer detector exploiting
  compressed sampling with a quadrature analog-to-information converter,''
  \emph{{IEEE} J. Solid-State Circuits}, vol.~50, no.~12, pp. 3047--3064, Dec.
  2015.

\bibitem{RFICYaz2016}
------, ``A compressed-sampling time-segmented quadrature analog-to-information
  converter for wideband rapid detection of up to 6 interferers with adaptive
  thresholding,'' in \emph{{IEEE} Radio Frequency Integrated Circuits
  Symposium}, May 2016, pp. 282--285.

\bibitem{eldar_cs_book}
Y.~C. Eldar and G.~Kutyniok, \emph{Compressed Sensing: Theory and
  Applications}.\hskip 1em plus 0.5em minus 0.4em\relax Cambridge University
  Press, 2012.

\bibitem{Haque2015}
T.~Haque and et~al., ``Theory and design of a quadrature analog-to-information
  converter for energy-efficient wideband spectrum sensing,'' \emph{{IEEE}
  Trans. Circuits Syst. {I}}, vol.~62, no.~2, pp. 527--535, Feb. 2015.

\bibitem{5711005}
M.~Saberi, R.~Lotfi, K.~Mafinezhad, and W.~A. Serdijn, ``Analysis of power
  consumption and linearity in capacitive digital-to-analog converters used in
  successive approximation adcs,'' \emph{IEEE Transactions on Circuits and
  Systems I: Regular Papers}, vol.~58, no.~8, pp. 1736--1748, 2011.

\bibitem{fixed_w_jssc13}
W.~Tang, A.~Osman, D.~Kim, B.~Goldstein, C.~Huang, B.~Martini, V.~A. Pieribone,
  and E.~Culurciello, ``Continuous time level crossing sampling adc for
  bio-potential recording systems,'' \emph{IEEE Transactions on Circuits and
  Systems I: Regular Papers}, vol.~60, no.~6, pp. 1407--1418, 2013.

\bibitem{szyduczynski2023time}
J.~Szyduczy{\'n}ski, D.~Ko{\'s}cielnik, and M.~Mi{\'s}kowicz, ``Time-to-digital
  conversion techniques: {A} survey of recent developments,''
  \emph{Measurement}, p. 112762, 2023.

\bibitem{vandePlassche2003}
\BIBentryALTinterwordspacing
R.~van~de Plassche, \emph{Sample-and-hold amplifiers}.\hskip 1em plus 0.5em
  minus 0.4em\relax Boston, MA: Springer US, 2003, pp. 313--347. [Online].
  Available: \url{https://doi.org/10.1007/978-1-4757-3768-4_7}
\BIBentrySTDinterwordspacing

\bibitem{wake-up_jssc21}
Z.~Wang, H.~Zhang, Y.~Zhang, L.~Shen, J.~Ru, H.~Fan, Z.~Tan, Y.~Wang, L.~Ye,
  and R.~Huang, ``A software-defined always-on system with 57–75-nw wake-up
  function using asynchronous clock-free pipelined event-driven architecture
  and time-shielding level-crossing adc,'' \emph{IEEE Journal of Solid-State
  Circuits}, vol.~56, no.~9, pp. 2804--2816, 2021.

\bibitem{Shlezinger_Entropy21}
N.~Shlezinger and Y.~C. Eldar, ``Deep task-based quantization,''
  \emph{Entropy}, vol.~23, no. 104, pp. 1--18, Jan. 2021.

\bibitem{Shlezinger_Tsp19Asym}
N.~Shlezinger, Y.~C. Eldar, and M.~R.~D. Rodrigues, ``Asymptotic task-based
  quantization with application to massive {MIMO},'' \emph{IEEE Trans. Signal
  Process.}, vol.~67, no.~15, pp. 3995--4012, Aug. 2019.

\bibitem{zirtiloglu_ts_mimo}
T.~Zirtiloglu, N.~Shlezinger, Y.~C. Eldar, and R.~Tugce~Yazicigil,
  ``Power-efficient hybrid mimo receiver with task-specific beamforming using
  low-resolution adcs,'' in \emph{ICASSP 2022 - 2022 IEEE International
  Conference on Acoustics, Speech and Signal Processing (ICASSP)}, 2022, pp.
  5338--5342.

\end{thebibliography}
\end{document}